\pdfoutput=1
\RequirePackage{ifpdf}
\ifpdf 
\documentclass[pdftex]{sigma}
\else
\documentclass{sigma}
\fi

\newtheorem{prop}{Proposition}[section]

\newtheorem{lem}{Lemma}[section]

\DeclareMathOperator{\tr}{tr}

\def\sk#1{\left(#1\right)}

\def\Uq#1{U_q(\widehat{\mathfrak{gl}}_{#1})}
\def\skk#1{\left[#1\right]}

\numberwithin{equation}{section}
\begin{document}

\allowdisplaybreaks

\renewcommand{\thefootnote}{$\star$}

\renewcommand{\PaperNumber}{058}

\FirstPageHeading

\ShortArticleName{Bethe Vectors of Quantum Integrable Models with GL(3) Trigonometric $R$-Matrix}

\ArticleName{Bethe Vectors of Quantum Integrable Models
\\
with GL(3) Trigonometric $\boldsymbol{R}$-Matrix\footnote{This paper is a~contribution to the Special Issue
in honor of Anatol Kirillov and Tetsuji Miwa.
The full collection is available at \href{http://www.emis.de/journals/SIGMA/InfiniteAnalysis2013.html}
{http://www.emis.de/journals/SIGMA/Inf\/initeAnalysis2013.html}}}

\Author{Samuel BELLIARD~$^{\dag^1}\!$, Stanislav PAKULIAK~$^{\dag^2\dag^3\dag^4}\!$,
Eric~RAGOUCY~$^{\dag^5}\!$, Nikita A.~SLAVNOV~$^{\dag^6}$} \AuthorNameForHeading{S.~Belliard, S.~Pakuliak,
E.~Ragoucy and N.A.~Slavnov}

\Address{$^{\dag^1}$~Universit\'e Montpellier 2, Laboratoire Charles Coulomb,
\\
\hphantom{$^{\dag^1}$}~UMR 5221, F-34095 Montpellier, France}
\EmailDD{\href{mailto:samuel.belliard@univ-montp2.fr}{samuel.belliard@univ-montp2.fr}}

\Address{$^{\dag^2}$~Laboratory of Theoretical Physics, JINR, 141980 Dubna, Moscow reg., Russia}
\EmailDD{\href{mailto:pakuliak@theor.jinr.ru}{pakuliak@theor.jinr.ru}}
\Address{$^{\dag^3}$~Moscow Institute of Physics and Technology, 141700 Dolgoprudny, Moscow reg., Russia}
\Address{$^{\dag^4}$~Institute of Theoretical and Experimental Physics, 117259 Moscow, Russia}

\Address{$^{\dag^5}$~Laboratoire de Physique Th\'eorique LAPTH, CNRS and Universit\'e de Savoie,
\\
\hphantom{$^{\dag^5}$}~BP 110, 74941 Annecy-le-Vieux Cedex, France}
\EmailDD{\href{mailto:eric.ragoucy@lapth.cnrs.fr}{eric.ragoucy@lapth.cnrs.fr}}

\Address{$^{\dag^6}$~Steklov Mathematical Institute, Moscow, Russia}
\EmailDD{\href{mailto:nslavnov@mi.ras.ru}{nslavnov@mi.ras.ru}}

\ArticleDates{Received May 27, 2013, in f\/inal form September 27, 2013; Published online October 07, 2013}

\Abstract{We study quantum integrable models with ${\rm GL}(3)$ trigonometric $R$-matrix and solvable by
the nested algebraic Bethe ansatz.
Using the presentation of the universal Bethe vectors in terms of projections of products of the currents
of the quantum af\/f\/ine algebra~$\Uq{3}$ onto intersections of dif\/ferent types of Borel subalgebras, we
prove that the set of the nested Bethe vectors is closed under the action of the elements of the monodromy
matrix.}


\Classification{81R50; 17B80}

\section{Introduction}

\renewcommand{\thefootnote}{\arabic{footnote}} \setcounter{footnote}{0}

We consider a~quantum integrable model def\/ined by the monodromy matrix $T(u)$ with matrix elements
$T_{ij}(u)$, $i,j=1,2,3$, which satisf\/ies the commutation relation
\begin{gather}
\label{RTT}
{\rm R}(u,v)\cdot(T(u)\otimes\mathbf{1})\cdot(\mathbf{1}\otimes T(v))=(\mathbf{1}
\otimes T(v))\cdot(T(u)\otimes\mathbf{1})\cdot{\rm R}(u,v),
\end{gather}
with the $\Uq{3}$ trigonometric quantum $R$-matrix
\begin{gather}
{\rm R}(u,v)={\sf f}(u,v)\sum_{1\leq i\leq3}{\sf E}_{ii}\otimes{\sf E}_{ii}+\sum_{1\leq i<j\leq3}
({\sf E}_{ii}\otimes{\sf E}_{jj}+{\sf E}_{jj}\otimes{\sf E}_{ii})
\nonumber
\\
\phantom{{\rm R}(u,v)=}{}
+\sum_{1\leq i<j\leq3}\big(u{\sf g}(u,v){\sf E}_{ij}\otimes{\sf E}_{ji}+v{\sf g}(u,v){\sf E}_{ji}
\otimes{\sf E}_{ij}\big).
\label{UqglN-R}
\end{gather}
Here the rational functions ${\sf f}(u,v)$ and ${\sf g}(u,v)$ are
\begin{gather*}
{\sf f}(u,v)=\frac{qu-q^{-1}v}{u-v},
\qquad
{\sf g}(u,v)=\frac{\big(q-q^{-1}\big)}{u-v},
\end{gather*}
and $({\sf E}_{ij})_{lk}=\delta_{il}\delta_{jk}$, $i,j,l,k=1,2,3$ are $3\times3$ matrices with unit in the
intersection of the $i$th row and the $j$th column and zero matrix elements elsewhere.
The $R$-matrix~\eqref{UqglN-R} is called `trigonometric' because its classical limit gives the classical
trigonometric $r$-matrix~\cite{BelDri82}.
The trigonometric $R$-matrix~\eqref{UqglN-R} is written in multiplicative variables and depends actually on
the ratio $u/v$ of these multiplicative parameters.

Due to the commutation relation~\eqref{RTT} the transfer matrix
$t(u)=\tr T(u)=T_{11}(u)+T_{22}(u)+T_{33}(u)$ generates a~set of commuting integrals of motion and the f\/irst
step of the algebraic Bethe ansatz~\cite{FadST79} is the construction of the set of eigenstates for these
commuting operators in terms of the monodromy matrix entries.
We assume that these matrix elements act in a~quantum space $V$ and that this space possesses a~vector
$|0\rangle\in V$ such that
\begin{gather*}
T_{ij}(u)|0\rangle=0,
\quad
i>j,
\qquad
T_{ii}(u)|0\rangle=\lambda_i(u)|0\rangle,
\qquad
\lambda_i(u)\in{\mathbb C}\big[\big[u,u^{-1}\big]\big].
\end{gather*}

The eigenstates ${\mathbb{B}}^{a,b}(\bar u;\bar v)$ of the transfer matrix $t(u)$ in quantum integrable
models with ${\rm GL}(3)$ trigonometric $R$-matrix depend on two sets of variables
\begin{gather*}
\bar u=\left\{u_1,\ldots,u_a\right\},
\qquad
\bar v=\left\{v_1,\ldots,v_b\right\},
\end{gather*}
which are called the Bethe parameters.
These eigenstates can be constructed in the framework of the nested Bethe ansatz method formulated
in~\cite{KulRes83} and are given by certain polynomials in the monodromy matrix elements $T_{12}(u)$,
$T_{23}(u)$, $T_{13}(u)$ with rational coef\/f\/icients depending on the Bethe parameters.

In pioneer papers on nested Bethe ansatz~\cite{KulRes81,KulRes82, KulRes83} no explicit formulae for the
Bethe vectors were obtained.
The method, in its original formulation, allows one to get the Bethe equations by requiring that the Bethe
vectors are eigenstates of the transfer matrix.
Nevertheless, even when the Bethe parameters are free and do not satisfy any restrictions, the structure of
the Bethe vectors (sometimes such Bethe vectors are called of\/f-shell) is rather complicated.
More explicit formulae for the of\/f-shell nested Bethe vectors were obtained in~\cite{VT} in the theory of
solutions of the quantum Knizhnik--Zamolodchikov equation.
The Bethe vectors were given by certain traces over auxiliary spaces of the products of the monodromy
matrices and $R$-matrices.
This presentation allows one to investigate the structure of the nested of\/f-shell Bethe vectors and to
obtain the explicit formulae for the nested Bethe vectors when the space $V$ becomes a~tensor product of
evaluation representations of the Yangian and of the positive Borel subalgebra of the quantum af\/f\/ine
algebra $\Uq{N}$~\cite{VTcom}.

Explicit expressions for the of\/f-shell nested Bethe vectors in the ${\rm GL}(N)$ quantum integrable
models in terms of the monodromy matrix elements were obtained in the papers~\cite{KP-GLN,KPT,OPS}, where
the realization of these vectors in terms of the current generators of the quantum af\/f\/ine algebra
$\Uq{N}$~\cite{EKhP} was used.
This realization uses the notion of projections onto intersections of dif\/ferent types of Borel
subalgebras in the quantum af\/f\/ine algebras introduced f\/irstly in~\cite{ER}.
Of course, it also uses the isomorphism between the current~\cite{D88} and the $L$-operator formulations of
the quantum af\/f\/ine algebras~\cite{RS} investigated in~\cite{DF}.

Quite analogously one can construct dual of\/f-shell Bethe vectors ${\mathbb{C}}^{a,b}(\bar u;\bar v)$
def\/ined in the dual space $V^*$ with the dual vacuum vector $\langle0|\in V^*$:
\begin{gather*}
\langle0|T_{ij}(u)=0,
\quad
i<j,
\qquad
\langle0 | T_{ii}(u)=\lambda_i(u)\langle0|.
\end{gather*}
They can be also explicitly written as polynomials in the monodromy matrix elements $T_{21}(u)$,
$T_{32}(u)$, $T_{31}(u)$ with rational coef\/f\/icients using the current realization of the quantum
af\/f\/ine algebra $\Uq{3}$~\cite{BPR}.

For the class of nested quantum integrable models where the inverse scattering problem can be solved and
local operators can be expressed in terms of the monodromy matrix elements~\cite{MaiTer00}, one can now
address the problem of calculation of the form factors and the correlation functions of local operators.
It was done in~\cite{KitMaiT99} for the quantum integrable models with ${\rm GL}(2)$ trigonometric
$R$-matrix, using determinant formulae for the scalar products of the Bethe vectors obtained in~\cite{Sl}.

To approach this problem one has to answer the following question.
Whether the action of the monodromy matrix elements onto nested of\/f-shell Bethe vectors produces linear
combinations of vectors with the same structure.
If this is true, then the problem of computing the form factors of local operators can be reduced to the
calculation of the scalar products between of\/f-shell and on-shell\footnote{These are the Bethe vectors
whose parameters satisfy the Bethe equations.} Bethe vectors.
Moreover, since right and left Bethe vectors are presented as linear combinations of products of the
monodromy matrix elements, the calculation of these scalar products itself can be also reduced to the
application of the action formulae of the monodromy matrix elements onto Bethe vectors.

The goal of this paper is to give a~positive answer to this question and to present and prove the explicit
formulae for such an action.
We should say that in case of quantum integrable models with ${\rm GL}(2)$ $R$-matrix, the question about
the action formulae is almost trivial, since the right and left of\/f-shell Bethe vectors in this case are
given by the product of the monodromy matrix elements $T_{12}(u)$ and $T_{21}(u)$ respectively.
These action formulae can be easily extracted from the $RTT$ relation~\eqref{RTT} for the monodromy
operators.
In higher-rank systems, due to the nontrivial structure of the nested Bethe vectors, the application of the
$RTT$ relations for the calculation of the action formulae becomes a~very complicated combinatorial problem.
In the following, to solve it, we will use the presentation of the nested of\/f-shell Bethe vectors in
terms of the current generators of the quantum af\/f\/ine algebra $\Uq{3}$ and the relation between the
monodromy matrix elements and the current generators given by the Gauss decomposition.

\section[Quantum af\/f\/ine algebra $\Uq{3}$]{Quantum af\/f\/ine algebra $\boldsymbol{\Uq{3}}$}

In order to reach the goal of the paper, rather than working with a~specif\/ic quantum integrable model
whose monodromy matrix satisf\/ies the commutation relations~\eqref{RTT}, we deal with a~more abstract
situation.
We consider the universal monodromy matrix which coincides with the $L$-operator of the positive Borel
subalgebra of the quantum af\/f\/ine algebra $\Uq{3}$.
There exists an isomorphism~\cite{DF} between the $L$-operators~\cite{RS} and the current~\cite{D88}
formulations of this algebra.
The expression of the universal Bethe vectors in terms of the current generators was computed
in~\cite{KP-GLN}, see also equations~\eqref{rbv2},~\eqref{lbv2} below.
Using these data, we will calculate the action of the monodromy matrix elements onto these Bethe vectors
using essentially the commutations relations of the algebra $\Uq{3}$ in the current realization.
The aim of this section is to introduce these algebraic objects.

\subsection[Two realizations of $\Uq{3}$]{Two realizations of $\boldsymbol{\Uq{3}}$}

The quantum af\/f\/ine algebra $\Uq{3}$ is an associative algebra with unit.
In the $L$-operator formulation~\cite{RS} it is generated by the modes ${L}^\pm_{ij}[n]$, $i,j=1,2,3$,
$n\geq0$ such that
\begin{gather}
\label{zeromode}
{L}^+_{ji}[0]={L}^-_{ij}[0]=0,
\qquad
1\leq i<j\leq3.
\end{gather}
These modes can be gathered into the generating series\footnote{There is also one relation for the zero modes
of the diagonal matrix elements of $L$-operators ${L}^+_{jj}[0]{L}^-_{jj}[0]=1$, $j=1,2,3$, which is not
important for our considerations.}
\begin{gather}
\label{Lop}
{L}^\pm(u)=\sum_{n\geq0}\sum_{i,j=1}^3{\sf E}_{ij}\otimes{L}^\pm_{ij}[n]u^{\mp n}
\in\text{End}\big({\mathbb C}^3\big)\otimes U_q(\mathfrak{b}_\pm),
\end{gather}
where $U_q(\mathfrak{b}_\pm)\subset \Uq{3}$ are the positive and negative Borel subalgebras of the quantum
af\/f\/ine algebra $\Uq{3}$.
These generating series can be called universal monodromy matrices since they satisfy the same
as~\eqref{RTT} commutation relation
\begin{gather}
\label{YB}
{\rm R}(u,v)\cdot\big({L}^\mu(u)\otimes\mathbf{1}\big)\cdot(\mathbf{1}\otimes{L}^\nu(v))
=(\mathbf{1}\otimes{L}^\nu(v))\cdot({L}^\mu(u)\otimes\mathbf{1})\cdot{\rm R}(u,v),
\end{gather}
where $\mu,\nu=\pm$.

The quantum af\/f\/ine algebra $\Uq{3}$ is a~Hopf algebra and the Borel subalgebras generated by the modes
of the $L$-operators ${L}^\pm(u)$ are Hopf subalgebras for the standard coproduct
\begin{gather*}
\Delta\sk{{L}_{ij}^\pm(u)}=\sum_{k=1}^3{L}^\pm_{kj}(u)\otimes{L}^\pm_{ik}(u).
\end{gather*}

In what follows we will need another realization of the same algebra, the so-called current realization of
the quantum af\/f\/ine algebra $\Uq{3}$ given in~\cite{D88}.
To relate the current and $L$-operator realizations of the same algebra we introduce, according
to~\cite{DF}, the Gauss decomposition of the $L$-operator
\begin{gather}
\label{Gaudec}
{L}^\pm(u)=\sk{
\begin{matrix}
1&{\rm F}^\pm_{21}(u)&{\rm F}^\pm_{31}(u)
\vspace{1mm}\\
0&1&{\rm F}^\pm_{32}(u)
\\
0&0&1
\end{matrix}
}\sk{
\begin{matrix}
{k}^\pm_1(u)&0&0
\\
0&{k}^\pm_2(u)&0
\\
0&0&{k}^\pm_3(u)
\end{matrix}
}\sk{
\begin{matrix}
1&0&0
\\
{\rm E}^\pm_{12}(u)&1&0
\vspace{1mm}\\
{\rm E}^\pm_{13}(u)&{\rm E}^\pm_{23}(u)&1
\end{matrix}
},
\end{gather}
that is to say
\begin{gather}
L^{\pm}_{ab}(u)={\rm F}^{\pm}_{ba}(u){k}^+_{b}(u)+\sum_{b<m\leq3}{\rm F}^{\pm}_{ma}(u){k}^+_{m}(t){\rm E}
^{\pm}_{bm}(u),
\qquad
a<b,
\label{GF2}
\\
L^{\pm}_{bb}(u)={k}^\pm_{b}(u)+\sum_{b<m\leq3}{\rm F}^{\pm}_{mb}(u){k}^\pm_{m}(u){\rm E}^{\pm}_{bm}(u),
\label{GF26}
\\
L^{\pm}_{ab}(u)={k}^\pm_{a}(u){\rm E}^{\pm}_{ba}(u)+\sum_{a<m\leq3}{\rm F}^{\pm}_{ma}(u){k}^\pm_{m}
(u){\rm E}^{\pm}_{bm}(u),
\qquad
a>b.
\label{GE2}
\end{gather}

It was proved in the paper~\cite{DF} that, after substitution of the
decompositions~\eqref{GF2}--\eqref{GE2} into the commutation relations~\eqref{YB}, one can obtain for the
linear combinations of the Gauss coordinates
\begin{gather}
\label{DF2}
{F}_i(t)={\rm F}^{+}_{i+1\,i}(t)-{\rm F}^{-}_{i+1\,i}(t),
\qquad
{E}_i(t)={\rm E}^{+}_{i\,i+1}(t)-{\rm E}^{-}_{i\,i+1}(t)
\end{gather}
and ${k}^\pm_i(t)$ the following commutation relations:
\begin{gather}
\big(q^{-1}z-qw\big){E}_{i}(z){E}_{i}(w)={E}_{i}(w){E}_{i}(z)\big(qz-q^{-1}w\big),
\\
(z-w){E}_{i}(z){E}_{i+1}(w)={E}_{i+1}(w){E}_{i}(z)\big(q^{-1}z-qw\big),
 \\
{k}_i^\pm(z){E}_i(w)\left({k}_i^\pm(z)\right)^{-1}=\frac{z-w}{q^{-1}z-qw}{E}_i(w),
\label{kiE}
\\
{k}_{i+1}^\pm(z){E}_i(w)\left({k}_{i+1}^\pm(z)\right)^{-1}=\frac{z-w}{qz-q^{-1}w}{E}_i(w),
\label{ki1E}
\\
\label{kE}
{k}_i^\pm(z){E}_j(w)\left({k}_i^\pm(z)\right)^{-1}={E}_j(w),
\qquad
{\rm if}
\quad
i\not=j,j+1,
\\[1.44mm]
\label{FiFi}
\big(qz-q^{-1}w\big){F}_{i}(z){F}_{i}(w)={F}_{i}(w){F}_{i}(z)\big(q^{-1}z-qw\big),
\\[1.44mm]
\label{FiFi1}
\big(q^{-1}z-qw\big){F}_{i}(z){F}_{i+1}(w)={F}_{i+1}(w){F}_{i}(z)(z-w),
\\
\label{kiF}
{k}_i^\pm(z){F}_i(w)\left({k}_i^\pm(z)\right)^{-1}=\frac{q^{-1}z-qw}{z-w}{F}_i(w),
\\
\label{ki1F}
{k}_{i+1}^\pm(z){F}_i(w)\left({k}_{i+1}^\pm(z)\right)^{-1}=\frac{qz-q^{-1}w}{z-w}{F}_i(w),
\\
{k}_i^\pm(z){F}_j(w)\left({k}_i^\pm(z)\right)^{-1}={F}_j(w),
\qquad
{\rm if}
\quad
i\not=j,j+1,
\\
\label{EF}
[{E}_{i}(z),{F}_{j}(w)]=\delta_{{i},{j}}\delta(z/w)\big(q-q^{-1}\big)\left({k}^+_{i}(z)/{k}^+_{i+1}
(z)-{k}^-_{i}(w)/{k}^-_{i+1}(w)\right),
\end{gather}
plus the Serre relations for the currents ${E}_{i}(z)$ and ${F}_{i}(z)$ which are unimportant for this
paper.

The commutation relations for the algebra $\Uq{3}$, given in terms of the currents, should be considered as
formal series identities describing the inf\/inite set of relations between the modes of these currents.
The symbol $\delta(z)$ entering these relations is the formal series $\sum\limits_{n\in{\mathbb Z}} z^n$.

For any series $G(t)=\sum\limits_{m\in{\mathbb Z}}G[m]t^{-m}$ we denote
$G(t)^{(+)}=\sum\limits_{m>0}G[m]t^{-m}$, and $G(t)^{(-)}=-\sum\limits_{ m\leq 0} G[m]  t^{-m}$.
Using this notation the
Ding--Frenkel formulae~\eqref{DF2} can be inverted
\begin{gather}
\label{DFinverse}
{\rm F}^{\pm}_{i+1\,i}(z)=z\left(z^{-1}{F}_i(z)\right)^{(\pm)},
\qquad
{\rm E}^{\pm}_{i\,i+1}(z)={E}_i(z)^{(\pm)}.
\end{gather}

\subsection{Dif\/ferent type Borel subalgebras and ordering of current generators}
\label{subal}

The isomorphism between the $L$-operator~\cite{RS} and the current~\cite{D88} formulations of the quantum
af\/f\/ine algebra, proved in~\cite{DF}, allows one to express the modes of the $L$-operators through the
modes of the currents and vice versa using the initial relation~\eqref{zeromode} and the
formulae~\eqref{GF2}--\eqref{GE2}.
On the other hand, it was proved in~\cite{KT} that the current generators for the quantum af\/f\/ine
algebras form the part of the Cartan--Weyl basis in these algebras.

There exists a~natural ordering in the Cartan--Weyl basis.
If the generator $e_\gamma$ corresponds to a~positive root $\gamma=\alpha+\beta$, where $\alpha$ and
$\beta$ are roots, then these generators are ordered either in a~way $e_\alpha\prec e_\gamma\prec e_\beta$
or in the way $e_\beta\prec e_\gamma\prec e_\alpha$.
An important property of the Cartan--Weyl basis of a~Borel subalgebra of the quantum algebras is that the
$q$-commutator of any two generators from this subalgebra, say $e_\alpha$ and $e_\beta$, is a~linear
combination of monomials containing only the products of generator $e_{\gamma_i}$ which are `between'
$e_\alpha$ and $e_\beta$:
\begin{gather*}
e_\alpha\prec e_{\gamma_i}\prec e_\beta
\qquad
\text{or} 
\qquad
e_\alpha\succ e_{\gamma_i}\succ e_\beta.
\end{gather*}
This property of the Cartan--Weyl basis allows one to describe easily the subalgebras in the quantum
af\/f\/ine algebras.
For instance, in the example above all generators corresponding to the roots $\alpha$, $\gamma_i$, $\beta$ form
a~subalgebra by def\/inition.
The standard positive Borel subalgebra in~$\Uq{3}$ generated by the modes of $L$-operators~\eqref{Lop} is
formed by the Cartan--Weyl ge\-ne\-rators which are `between' the af\/f\/ine root generator $e_{\alpha_0}$ and
non-af\/f\/ine negative simple roots generators~$e_{-\alpha_1}$ and~$e_{-\alpha_2}$.
Respectively, the negative Borel subalgebra is formed by the generators which are `between'
$e_{\alpha_1}$, $e_{\alpha_2}$ and $e_{-\alpha_0}$.

The ordering on the Borel subalgebra can be extended to the ordering of the whole set of Cartan--Weyl
generators corresponding to the positive and negative roots such that the same ordering property is valid.
This ordering is called `circular' or `convex' and it allows one to order arbitrary monomials in the whole
algebra~\cite{EKhP}.

We consider two types of Borel subalgebras of the algebra $\Uq{3}$.
Standard positive and negative Borel subalgebras $U_q(\mathfrak{b}^\pm)\subset \Uq{3} $ are generated by
the modes of the $L$-operators $L^{(\pm)}(u)$ respectively.
For the generators in these subalgebras we can use the modes of the Gauss
coordinates~\eqref{GF2}--\eqref{GE2} ${\rm E}^\pm_{i\,i+1}(u)$, ${\rm F}^\pm_{i+1\,i}(u)$,
${k}^\pm_{j}(u)$, $i=1,2$, $j=1,2,3$.

Another type of Borel subalgebras is related to the current realizations of $\Uq{3}$ given in the
previous subsection.
The Borel subalgebra $U_F\subset \Uq{3}$ is generated by modes of the currents $F_i[n]$, $k^+_j[m]$,
$i=1,2$, $j=1,2,3$, $n\in{\mathbb Z}$ and $m\geq0$.
The Borel subalgebra $U_E\subset \Uq{3}$ is generated by the modes of the currents $E_i[n]$, $k^-_j[-m]$,
$i=1,2$, $j=1,2,3$, $n\in{\mathbb Z}$ and $m\geq0$.
We will consider also a~subalgebras $U'_F=U_F\setminus\{k^+_j[0]\}$ and
$U'_E=U_E\setminus\{k^-_j[0]\}$.\footnote{In order to obtain the quantum af\/f\/ine algebra $\Uq{3}$ in the
framework of the quantum double construction~\cite{D88} one has to impose the relation
$k^+_j[0]k^-_j[0]=1$, $j=1,2,3$.}

Further, we will be interested in the intersections,
\begin{gather*}
{U}_F^-={U}'_F\cap U_q(\mathfrak{b}^-),
\qquad
{U}_F^+={U}_F\cap U_q(\mathfrak{b}^+),
\\
{U}_E^-={U}_E\cap U_q(\mathfrak{b}^-),
\qquad
{U}_E^+={U}'_E\cap U_q(\mathfrak{b}^+),
\end{gather*}
and will describe properties of projections to these intersections.
We call $U_F$ and $U_E$ \emph{the current Borel subalgebras}.
Let $U_f\subset U_F$ and $U_e\subset U_E$ be the subalgebras of the current Borel subalgebras generated by
the modes of the currents $F_i[n]$ and $E_i[n]$, $i=1,2$, $n\in{\mathbb Z}$ only.
In what follows we will use the subalgebras $U^+_f\subset U_f$ and $U^+_e\subset U_e$ def\/ined by the intersections
\begin{gather*}
U_f^+=U^+_F\cap U_f
\qquad
U_e^+=U^+_E\cap U_e.
\end{gather*}
Let $U^\pm_k$ be subalgebras in $\Uq{3}$ generated by the modes of the Cartan currents $k^\pm_j(u)$.

We f\/ix a~`circular' ordering `$\prec$' on the generators of $\overline{U}_q(\mathfrak{gl}_3)$
(see~\cite{EKhP}), such that:
\begin{gather}
\label{circular}
\cdots\prec U^-_k\prec U^-_f\prec U^+_f\prec U^+_k\prec U^+_e\prec U^-_e\prec U^-_k\prec\cdots.
\end{gather}

\begin{figure}[t]
\centering
\includegraphics{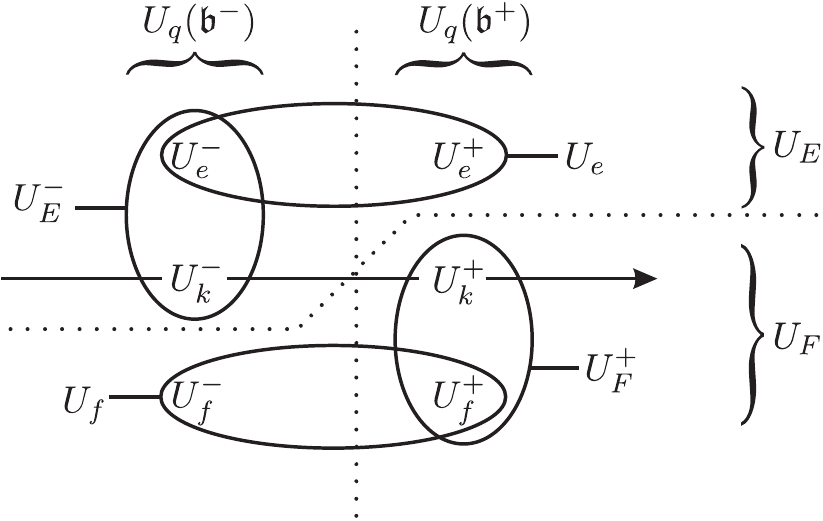}

\caption{Subalgebras of $\overline{U}_q(\mathfrak{gl}_3)$.
The vertical dotted line separates the standard Borel subalgebras.
The horizontal dotted line separates the current Borel subalgebras.
The horizontal solid axis indicates the increasing of the current generators modes.
Ovals denote dif\/ferent subalgebras in the $\overline{U}_q(\mathfrak{gl}_3)$ standard and current Borel
subalgebras.}
\label{fig:subalg}
\end{figure}

The ordering of the subalgebras described above can be pictured in the Fig.~\ref{fig:subalg} in the
anti-clockwise direction.

We will call an element $W\in\overline{U}_q(\mathfrak{gl}_3)$ \emph{normal ordered} and denote it as
${:}W{:}$ if it is presented as linear combinations of products $W_1\cdot W_2\cdot W_3\cdot W_4\cdot
W_5\cdot W_6$ such that
\begin{gather*}
W_1\in U^-_f,
\qquad
W_2\in U^+_f,
\qquad
W_3\in U^+_k,
\qquad
W_4\in U^+_e,
\qquad
W_5\in U^-_e,
\qquad
W_6\in U^-_k.
\end{gather*}

We may consider the standard Borel subalgebras as ordered with respect to the circular
ordering~\eqref{circular}:
\begin{gather*}
U_q\big(\mathfrak{b}^-\big)=U^-_e\cdot U^-_k\cdot U^-_f,
\qquad
U_q\big(\mathfrak{b}^+\big)=U^+_f\cdot U^+_k\cdot U^+_e.
\end{gather*}
An analogous statement is valid for the current Borel subalgebras:
\begin{gather*}
U_F=U^-_f\cdot U^+_f\cdot U^+_k,
\qquad
U_E=U^+_e\cdot U^-_e\cdot U^-_k.
\end{gather*}

Let us note that the matrix elements in the universal monodromy matrix ${L}^+(u)$ given by the
formulae~\eqref{GF2}--\eqref{GE2} are normal ordered, i.e.
{}${:}{L}^+(u){:} ={L}^+(u)$.
The problem which we address in this paper, namely the calculation of the action of the monodromy matrix
elements onto of\/f-shell Bethe vectors, can be reformulated in the following way.
We should put the product of these elements and the element ${P}^+_f\left({F}_{2}(v_b)\cdots
{F}_{2}(v_1)\cdot {F}_{1}(u_a)\cdots {F}_{1}(u_1)\right)\in U^+_f$ into its normal order form, modulo terms
which annihilate the right vacuum vector $|0\rangle$.
Using the Gauss decompositions~\eqref{GF2}--\eqref{GE2}, it could be reduced to the commutation of the
Gauss coordina\-tes~$E^+_{ij}(u)$ with the element ${P}^+_f\left({F}_{2}(v_b)\cdots {F}_{2}(v_1)\cdot
{F}_{1}(u_a)\cdots {F}_{1}(u_1)\right)$.
However, this way of doing the normal ordering is almost equivalent to the use of the $RTT$ commutation
relations and is far too complicated to be useful for our purpose.

In fact, in this paper, we will employe a~dif\/ferent and more ef\/f\/icient strategy: we will use the
method of projections introduced in~\cite{ER} and exploited in a~series of papers (see~\cite{KP-GLN} and
references therein) to relate the of\/f-shell Bethe vectors with the current realization of the quantum
af\/f\/ine algebras.
We refer the reader to the above mentioned papers to f\/ind a~complete theory of the projections onto
intersections of the dif\/ferent types of Borel subalgebras.
Here, we will give only some short def\/initions on projections.
In order to do this, we need to equip the algebra $\Uq{3}$ together with its decomposition into current
Borel subalgebras by the \emph{current} Hopf structure
\begin{gather}
\Delta^{(D)}\sk{{E}_i(z)}={E}_i(z)\otimes1+{k}^-_{i}(z)\sk{{k}^-_{i+1}(z)}^{-1}\otimes{E}_i(z),
\nonumber
\\
\Delta^{(D)}\sk{{F}_i(z)}=1\otimes{F}_i(z)+{F}_i(z)\otimes{k}^+_{i}(z)\sk{{k}^+_{i+1}(z)}^{-1},
\nonumber
\\
\Delta^{(D)}\sk{{k}^\pm_i(z)}={k}^\pm_i(z)\otimes{k}^\pm_{i}(z).
\label{gln-cm}
\end{gather}

According to the general theory~\cite{EKhP} we introduce the projection operators
\begin{gather*}
{P}^\pm_{f}:U_F\subset\Uq{3}\to U_F^\pm,
\qquad
{P}^\pm_e:U_E\subset\Uq{3}\to U_E^\pm.
\end{gather*}
They are respectively def\/ined by the prescriptions
\begin{gather}
\label{pgln-d}
{P}^+_f(f_-f_+)=\varepsilon(f_-)f_+,
\qquad
{P}^-_f(f_-f_+)=f_-\varepsilon(f_+),
\qquad
\forall\,f_-\in{U}_F^-,
\quad
\forall\,f_+\in{U}_F^+,
\\
\label{pgln2-d}
{P}^+_e(e_+e_-)=e_+\varepsilon(e_-),
\qquad
{P}^-_e(e_-e_+)=\varepsilon(e_+)e_-,
\qquad
\forall\,e_-\in{U}_E^-,
\quad
\forall\,e_+\in{U}_E^+,
\end{gather}
where the counit map $\varepsilon:\Uq{3}\to{\mathbb C}$ is def\/ined on current generators as follows
\begin{gather*}
\varepsilon(1)=\varepsilon\big(k_j^\pm(u)\big)=1,
\qquad
\varepsilon(E_i(u))=\varepsilon\sk{F_i(u)}=0.
\end{gather*}

Denote by ${\overline{U}}_F$ and ${\overline{U}}_E$ the extensions of the algebras $U_F$ and $U_E$ formed
by inf\/inite sums of monomials which are ordered products $a_{i_1}[n_1]\cdots a_{i_k}[n_k]$ with
$n_1\leq\cdots\leq n_k$, where $a_{i_l}[n_l]$ is either $F_{i_l}[n_l]$ or $k^+_{i_l}[n_l]$ and
$E_{i_l}[n_l]$ or $k^-_{i_l}[n_l]$, respectively.
It can be checked that
\begin{itemize}\itemsep=0pt
\item[(1)] the action of the projections~\eqref{pgln-d} can be extended to the algebra ${\overline U}_F$;
\item[(2)] for any $f\in {\overline U}_F$ with $\Delta^{(D)}(f)=\sum\limits_i f'_i\otimes f''_i$ we have
\begin{gather}
\label{hF-ord-a}
f=\sum_i{P}^-_f(f''_i)\cdot{P}^+_f(f'_i);
\end{gather}
\item[(3)] the action of the projections~\eqref{pgln2-d} can be extended to the algebra ${\overline U}_E$;
\item[(4)] for any $e\in {\overline U}_E$ with $\Delta^{(D)}(e)=\sum\limits_i e'_i\otimes e''_i$ we have
\begin{gather}
\label{hE-ord-a}
e=\sum_i{P}^+_e(e'_i)\cdot{P}^-_e(e''_i).
\end{gather}
\end{itemize}

The formulae~\eqref{hF-ord-a} and~\eqref{hE-ord-a} are the main technical tools to calculate the
projections of currents.
These formulae allow us to present a~product of currents in a~normal ordered form using projections and the
rather simple current Hopf structure~\eqref{gln-cm}.

The Ding--Frenkel isomorphism between $L$-operator and current realizations of the quantum af\/f\/ine
algebra $\Uq{N}$~\cite{DF} identif\/ies the Gauss coordinates and the full currents through
formulae~\eqref{DF2} and~\eqref{DFinverse}.
It is clear that the Gauss coordinates ${\rm F}^\pm_{i+1\, i}(u)={P}^\pm_{f}\sk{F_i(u)}$ and ${\rm
E}^\pm_{i\, i+1}={P}^\pm_e\sk{E_i(u)}$ are def\/ined by the corresponding projections of the full currents.
But there are also higher Gauss coordinates ${\rm F}^\pm_{ji}(u)$ and ${\rm E}^\pm_{ij}(u)$ for $j>i+1$ and
their relation to the currents was not established in~\cite{DF}.
In~\cite{KP-GLN}, special elements from the completed algebras $\overline{U}_F$ and $\overline{U}_E$ were
introduced such that their projections yield the corresponding higher Gauss coordinates.
These elements were called `composed' currents.
In the case of the quantum af\/f\/ine algebra $\Uq{3}$, there are only two composed currents
\begin{gather}
\label{com-cur}
F_{3,1}(u)\equiv\big(q-q^{-1}\big)F_1(u)F_2(u),
\qquad
E_{1,3}(u)\equiv\big(q-q^{-1}\big)E_2(u)E_1(u),
\end{gather}
such that
\begin{gather*}
{P}^+_f\sk{F_{3,1}(u)}=\big(q-q^{-1}\big){\rm F}^+_{31}(u),
\qquad
{P}^+_e\sk{E_{1,3}(u)}=\big(q-q^{-1}\big){\rm E}^+_{13}(u).
\end{gather*}

\section{Main results}

\subsection{Notations}

To save space and simplify presentation, we use the following convention for the products of the commuting
entries of the monodromy matrix $T_{ij}(w)$, vacuum eigenvalues $\lambda_i(w)$ and their ratios ${\sf
r}_k(w)=\lambda_k(w)/\lambda_2(w)$, $k=1,3$.
Namely, whenever such an operator or a~scalar function depends on a~set of variables (for instance,
$T_{ij}(\bar w)$, $\lambda_i(\bar u)$, ${\sf r}_k(\bar v)$), this means that we deal with the product of
the operators or scalar functions with respect to the corresponding set:
\begin{gather*}
T_{ij}(\bar w)=\prod_{w_k\in\bar w}T_{ij}(w_k);
\qquad
\lambda_2(\bar u)=\prod_{u_j\in\bar u}\lambda_2(u_j);
\qquad
{\sf r}_k(\bar v_\ell)=\prod_{\substack{v_j\in\bar v
\\
v_j\ne v_\ell}}{\sf r}_k(v_j).
\end{gather*}
A similar convention will be used for the products of functions ${\sf f}(u,v)$ and ${\sf g}(u,v)$
\begin{gather*}
{\sf f}(w_i,\bar w_i)=\prod_{\substack{w_j\in\bar w
\\
w_j\ne w_i}}{\sf f}(w_i,w_j);
\qquad
{\sf g}(\bar u,\bar v)=\prod_{u_j\in\bar u}\prod_{v_k\in\bar v}{\sf g}(u_j,v_k).
\end{gather*}
The notation $\bar v_\ell$ for an arbitrary set $\bar v$ means the set $\bar v\setminus \{v_\ell$\}.
We will also use the sets $\bar w_{<j}=\{w_1,...,w_{j-1}\}$ and $\bar w_{>j}=\bar w_j \setminus \bar
w_{<j}$ with obvious convention for the products.
Partitions of sets will be noted as $\bar u \Rightarrow \{\bar u_{\scriptscriptstyle \rm I},\, \bar
u_{\scriptscriptstyle \rm I\hspace{-1pt}I}\}$.

To simplify further formulae we will introduce a~special notation for product of non-com\-muting currents:
\begin{gather}
\label{For}
{\cal F}_1(\bar u)={F}_1(u_a){F}_1(u_{a-1})\cdots{F}_1(u_1),
\qquad
{\cal F}_2(\bar v)={F}_2(v_b)\cdots{F}_2(v_2){F}_2(v_1)
\end{gather}
and
\begin{gather}
{\cal F}_1(\bar u_j)={F}_1(u_a)\cdots{F}_1(u_{j+1}){F}_1(u_{j-1})\cdots{F}_1(u_1),
\nonumber
\\
{\cal F}_2(\bar v_i)={F}_2(v_b)\cdots{F}_2(v_{i+1}){F}_2(v_{i-1})\cdots{F}_2(v_1).
\label{For1}
\end{gather}
These notations are in accordance with the one used for commuting objects, except that now one needs to
specify the order as prescribed in~\eqref{For} and~\eqref{For1}.

In various formulae below the Izergin determinant ${\sf K}_k(\bar x|\bar y)$ appears~\cite{Ize87}.
It is def\/ined for two sets $\bar x$ and $\bar y$ of the same cardinality $\#\bar x=\#\bar y=k$:
\begin{gather}
\label{Izer}
{\sf K}_k(\bar x|\bar y)=\frac{\prod\limits_{1\leq i,j\leq k}(qx_i-q^{-1}y_j)}{\prod\limits_{1\leq i<j\leq k}
(x_i-x_j)(y_j-y_i)}\cdot\det\left[\frac{q-q^{-1}}{(x_i-y_j)(qx_i-q^{-1}y_j)}\right].
\end{gather}
Below we also use two modif\/ications of the Izergin determinant
\begin{gather}
\label{Mod-Izer}
{\sf K}^{(l)}_k(\bar x|\bar y)=\prod_{i=1}^kx_i\cdot{\sf K}_k(\bar x|\bar y),
\qquad
{\sf K}^{(r)}_k(\bar x|\bar y)=\prod_{i=1}^ky_i\cdot{\sf K}_k(\bar x|\bar y).
\end{gather}
Some properties of the Izergin determinant and its modif\/ications are gathered into Appendix~\ref{apI}.

\subsection{Explicit expression for Bethe vectors}

The right and left of\/f-shell Bethe vectors can be presented using the current realization of the quantum
af\/f\/ine algebra $\Uq{3}$~\cite{KP-GLN}
\begin{gather}
\label{rbv2}
{\mathbb{B}}^{a,b}(\bar u;\bar v)=\frac{\beta(\bar u|\bar v)}{{\sf f}(\bar v,\bar u)}{P}^+_f\left({F}_{2}
(v_b)\cdots{F}_{2}(v_1)\cdot{F}_{1}(u_a)\cdots{F}_{1}(u_1)\right)\cdot{\sf r}_3(\bar v)|0\rangle,
\\
\label{lbv2}
{\mathbb{C}}^{a,b}(\bar u;\bar v)=\frac{\beta(\bar u|\bar v)}{{\sf f}(\bar v,\bar u)}\langle0|{\sf r}
_3(\bar v){P}^+_e\left({E}_{1}(u_1)\cdots{E}_{1}(u_a)\cdot{E}_{2}(v_1)\cdots{E}_{2}(v_b)\right),
\end{gather}
where
\begin{gather*}
\beta(\bar u|\bar v)=\prod_{1\leq\ell<\ell'\leq a}{\sf f}(u_{\ell'},u_\ell)\prod_{1\leq\ell<\ell'\leq b}
{\sf f}(v_{\ell'},v_\ell),
\end{gather*}
and ${P}^+_f$ and ${P}^+_e$ are projections onto subalgebras of $\Uq{3}$ generated by the non-negative and
positive modes of the simple root currents ${F}_i(u)$ and ${E}_i(u)$, $i=1,2$, respectively.
These projections  onto subalgebras in the positive Borel subalgebra of $\Uq{3}$ were introduced
in~\cite{ER} and their detailed theory was developed in~\cite{EKhP}.
The formal def\/inition of these projections is given in the present paper through the
formulae~\eqref{pgln-d} and~\eqref{pgln2-d}.

In what follows we will consider the action of the universal monodromy matrix elements expressed in terms
of the Gauss coordinates~\eqref{Gaudec} or in terms of the current generators of the quantum af\/f\/ine
algebra $\Uq{3}$ onto universal of\/f-shell Bethe vectors~\eqref{rbv2}.
To obtain explicit formulae for this action we do not need to calculate the projection in~\eqref{rbv2}, but
use a~special presentation for this projection found in~\cite{FKPR} (see also~\eqref{us-P2} below).
Using this presentation we only need the commutation relations of the total currents which are much more
simple than the $RTT$-relations or the relations between Gauss coordinates.

Note that the function $\beta(\bar u|\bar v)$ removes all the poles and zeros which originate from the
product of currents of the same type, while the product of functions ${\sf f}(\bar v,\bar u)$ removes all
the poles which originate from the product of currents of dif\/ferent types.
Indeed, the product ${F}_i(u_2){F}_i(u_1)$ has a~simple pole at the point $u_1=q^2u_2$ and a~simple zero at
$u_1=u_2$, while the product ${F}_2(v){F}_1(u)$ has a~simple pole at the point $u=v$.
These `analytical' properties of the product of currents are determined by the commutation
relations~\eqref{FiFi}, \eqref{FiFi1} and were explained in details in the papers~\cite{KP-GLN,OPS} using
the notion of ordering of the current generators.

\subsection[Multiple action of $T_{ij}(\bar w)$ operators on Bethe vectors]{Multiple action of
$\boldsymbol{T_{ij}(\bar w)}$ operators on Bethe vectors}

Now we give the main result of this paper, namely a~complete list of the multiple actions of the operators
$T_{ij}(\bar w)$ onto the Bethe vectors $\mathbb{B}^{a,b}(\bar u;\bar v)$.
\begin{prop}
Throughout the proposition, we denote $\{\bar v,\bar w\}=\bar\xi$, $\{\bar u,\bar w\}=\bar\eta$ and $\#\bar w=n$.

The multiple actions of the $T_{ij}(\bar w)$ operators onto the Bethe vectors $\mathbb{B}^{a,b}(\bar u;\bar v)$
are given by:
\begin{itemize}\itemsep=0pt
\item Multiple action of $T_{13}$
      \begin{gather}
      \label{act13}
      T_{13}(\bar w)\mathbb{B}^{a,b}(\bar u;\bar v)=\lambda_2(\bar w)\,\mathbb{B}^{a+n,b+n}(\bar\eta;\bar\xi).
      \end{gather}
\item Multiple action of $T_{12}$
      \begin{gather}
      \label{act12}
      T_{12}(\bar w)\mathbb{B}^{a,b}(\bar u;\bar v)=\lambda_2(\bar w) \sum\frac{{\sf f}
      (\bar\xi_{\scriptscriptstyle\rm I\hspace{-1pt}I},\bar\xi_{\scriptscriptstyle\rm I})}{{\sf f}
      (\bar w,\bar\xi_{\scriptscriptstyle\rm I})}{\sf K}^{(r)}_n(\bar w|\bar\xi_{\scriptscriptstyle\rm I}
      )\,\mathbb{B}^{a+n,b}(\bar\eta;\bar\xi_{\scriptscriptstyle\rm I\hspace{-1pt}I}).
      \end{gather}
      The sum is taken over partitions of
      $\bar\xi\Rightarrow\{\bar\xi_{\scriptscriptstyle \rm I},
      \bar\xi_{\scriptscriptstyle \rm I\hspace{-1pt}I}\}$ with $\#\bar\xi_{\scriptscriptstyle \rm I}=n$.
\item Multiple action of $T_{23}$
      \begin{gather}
      \label{act23}
      T_{23}(\bar w)\mathbb{B}^{a,b}(\bar u;\bar v)=\lambda_2(\bar w) \sum\frac{{\sf f}
      (\bar\eta_{\scriptscriptstyle\rm I},\bar\eta_{\scriptscriptstyle\rm I\hspace{-1pt}I})}{{\sf f}
      (\bar\eta_{\scriptscriptstyle\rm I},\bar w)}{\sf K}^{(l)}_n(\bar\eta_{\scriptscriptstyle\rm I}
      |\bar w)\,\mathbb{B}^{a,b+n}(\bar\eta_{\scriptscriptstyle\rm I\hspace{-1pt}I};\bar\xi).
      \end{gather}
      The sum is taken over partitions of
      $\bar\eta\Rightarrow\{\bar\eta_{\scriptscriptstyle \rm I},
      \bar\eta_{\scriptscriptstyle \rm I\hspace{-1pt}I}\}$ with $\#\bar\eta_{\scriptscriptstyle \rm I}=n$.
\item Multiple action of $T_{22}$
      \begin{gather}
      \label{act22}
      T_{22}(\bar w)\mathbb{B}^{a,b}(\bar u;\bar v)=\lambda_2(\bar w) \sum\frac{{\sf f}
      (\bar\xi_{\scriptscriptstyle\rm I\hspace{-1pt}I},\bar\xi_{\scriptscriptstyle\rm I}){\sf f}
      (\bar\eta_{\scriptscriptstyle\rm I},\bar\eta_{\scriptscriptstyle\rm I\hspace{-1pt}I})}{{\sf f}
      (\bar w,\bar\xi_{\scriptscriptstyle\rm I}){\sf f}(\bar\eta_{\scriptscriptstyle\rm I},\bar w)}{\sf K}^{(r)}
      _n(\bar w|\bar\xi_{\scriptscriptstyle\rm I})\,{\sf K}^{(l)}_n(\bar\eta_{\scriptscriptstyle\rm I}
      |\bar w)\,\mathbb{B}^{a,b}(\bar\eta_{\scriptscriptstyle\rm I\hspace{-1pt}I}
      ;\bar\xi_{\scriptscriptstyle\rm I\hspace{-1pt}I}).
      \end{gather}
            The sum is taken over partitions of:
            $\bar\eta\Rightarrow\{\bar\eta_{\scriptscriptstyle \rm
      I},\bar\eta_{\scriptscriptstyle \rm I\hspace{-1pt}I}\}$ with $\#\bar\eta_{\scriptscriptstyle \rm I}=n$;
      $\bar\xi\Rightarrow\{\bar\xi_{\scriptscriptstyle \rm I},\bar\xi_{\scriptscriptstyle \rm
      I\hspace{-1pt}I}\}$ with $\#\bar\xi_{\scriptscriptstyle \rm I}=n$.

\item Multiple action of $T_{11}$
      \begin{gather}
      T_{11}(\bar w)\mathbb{B}^{a,b}(\bar u;\bar v)
      \nonumber
      \\
      \qquad
      {}=\lambda_2(\bar w) \sum\frac{{\sf r}
      _1(\bar\eta_{\scriptscriptstyle\rm I})}{{\sf f}(\bar\xi_{\scriptscriptstyle\rm I\hspace{-1pt}I}
      ,\bar\eta_{\scriptscriptstyle\rm I})}\,\frac{{\sf f}(\bar\xi_{\scriptscriptstyle\rm I\hspace{-1pt}I}
      ,\bar\xi_{\scriptscriptstyle\rm I}){\sf f}(\bar\eta_{\scriptscriptstyle\rm I\hspace{-1pt}I}
      ,\bar\eta_{\scriptscriptstyle\rm I})}{{\sf f}(\bar w,\bar\xi_{\scriptscriptstyle\rm I}){\sf f}
      (\bar\xi_{\scriptscriptstyle\rm I},\bar\eta_{\scriptscriptstyle\rm I})}{\sf K}^{(r)}
      _n(\bar w|\bar\xi_{\scriptscriptstyle\rm I})\,{\sf K}^{(r)}_n(\bar\xi_{\scriptscriptstyle\rm I}
      |\bar\eta_{\scriptscriptstyle\rm I})\,\mathbb{B}^{a,b}(\bar\eta_{\scriptscriptstyle\rm I\hspace{-1pt}I}
      ;\bar\xi_{\scriptscriptstyle\rm I\hspace{-1pt}I}).\label{act11}
      \end{gather}
      The sum is taken over partitions of:
      $\bar\eta\Rightarrow\{\bar\eta_{\scriptscriptstyle \rm
      I},\bar\eta_{\scriptscriptstyle \rm I\hspace{-1pt}I}\}$ with $\#\bar\eta_{\scriptscriptstyle \rm I}=n$;
      $\bar\xi\Rightarrow\{\bar\xi_{\scriptscriptstyle \rm I},\bar\xi_{\scriptscriptstyle \rm
      I\hspace{-1pt}I}\}$ with $\#\bar\xi_{\scriptscriptstyle \rm I}=n$.

\item Multiple action of $T_{33}$
      \begin{gather}
      T_{33}(\bar w)\mathbb{B}^{a,b}(\bar u;\bar v)
      \nonumber
      \\
      \qquad
     {} =\lambda_2(\bar w) \sum\frac{{\sf r}
      _3(\bar\xi_{\scriptscriptstyle\rm I})}{{\sf f}(\bar\xi_{\scriptscriptstyle\rm I}
      ,\bar\eta_{\scriptscriptstyle\rm I\hspace{-1pt}I})}\,\frac{{\sf f}(\bar\xi_{\scriptscriptstyle\rm I}
      ,\bar\xi_{\scriptscriptstyle\rm I\hspace{-1pt}I}){\sf f}(\bar\eta_{\scriptscriptstyle\rm I}
      ,\bar\eta_{\scriptscriptstyle\rm I\hspace{-1pt}I})}{{\sf f}(\bar\xi_{\scriptscriptstyle\rm I}
      ,\bar\eta_{\scriptscriptstyle\rm I}){\sf f}(\bar\eta_{\scriptscriptstyle\rm I},\bar w)}{\sf K}^{(l)}
      _n(\bar\eta_{\scriptscriptstyle\rm I}|\bar w)\,{\sf K}^{(l)}_n(\bar\xi_{\scriptscriptstyle\rm I}
      |\bar\eta_{\scriptscriptstyle\rm I})\,\mathbb{B}^{a,b}(\bar\eta_{\scriptscriptstyle\rm I\hspace{-1pt}I}
      ;\bar\xi_{\scriptscriptstyle\rm I\hspace{-1pt}I}).\label{act33}
      \end{gather}
The sum is taken over partitions of:
      $\bar\eta\Rightarrow\{\bar\eta_{\scriptscriptstyle \rm
      I},\bar\eta_{\scriptscriptstyle \rm I\hspace{-1pt}I}\}$ with $\#\bar\eta_{\scriptscriptstyle \rm I}=n$;
      $\bar\xi\Rightarrow\{\bar\xi_{\scriptscriptstyle \rm I},\bar\xi_{\scriptscriptstyle \rm
      I\hspace{-1pt}I}\}$ with $\#\bar\xi_{\scriptscriptstyle \rm I}=n$.

\item Multiple action of $T_{21}$
      \begin{gather}
      T_{21}(\bar w)\mathbb{B}^{a,b}(\bar u;\bar v)=\lambda_2(\bar w) \sum{\mathsf r}
      _1(\bar\eta_{\scriptscriptstyle\rm I})\,\frac{{\sf f}(\bar\eta_{\scriptscriptstyle\rm I\hspace{-1pt}I}
      ,\bar\eta_{\scriptscriptstyle\rm I}){\sf f}(\bar\eta_{\scriptscriptstyle\rm I\hspace{-1pt}I}
      ,\bar\eta_{\scriptscriptstyle\rm I\hspace{-1pt}I\hspace{-1pt}I}){\sf f}
      (\bar\eta_{\scriptscriptstyle\rm I\hspace{-1pt}I\hspace{-1pt}I},\bar\eta_{\scriptscriptstyle\rm I}){\sf f}
      (\bar\xi_{\scriptscriptstyle\rm I\hspace{-1pt}I},\bar\xi_{\scriptscriptstyle\rm I})}{{\sf f}
      (\bar\xi,\bar\eta_{\scriptscriptstyle\rm I}){\sf f}(\bar w,\bar\xi_{\scriptscriptstyle\rm I}){\sf f}
      (\bar\eta_{\scriptscriptstyle\rm I\hspace{-1pt}I},\bar w)}
      \nonumber
      \\
      \qquad
     {} \times{\sf K}^{(l)}_n(\bar\eta_{\scriptscriptstyle\rm I\hspace{-1pt}I}|\bar w)\,{\sf K}^{(r)}
      _n(\bar\xi_{\scriptscriptstyle\rm I}|\bar\eta_{\scriptscriptstyle\rm I})\,{\sf K}^{(r)}
      _n(\bar w|\bar\xi_{\scriptscriptstyle\rm I})\,\,\mathbb{B}^{a-n,b}
      (\bar\eta_{\scriptscriptstyle\rm I\hspace{-1pt}I\hspace{-1pt}I}
      ;\bar\xi_{\scriptscriptstyle\rm I\hspace{-1pt}I}).
      \label{act21}
      \end{gather}
 The sum is taken over partitions of:
      $\bar\eta\Rightarrow\{\bar\eta_{\scriptscriptstyle \rm I},
      \bar\eta_{\scriptscriptstyle \rm I\hspace{-1pt}I},
      \bar\eta_{\scriptscriptstyle \rm I\hspace{-1pt}I\hspace{-1pt}I}\}$
      with
      $\#\bar\eta_{\scriptscriptstyle \rm I}=\#\bar\eta_{\scriptscriptstyle \rm I\hspace{-1pt}I}=n$;
      $\bar\xi\Rightarrow\{\bar\xi_{\scriptscriptstyle \rm I},\bar\xi_{\scriptscriptstyle \rm
      I\hspace{-1pt}I}\}$ with $\#\bar\xi_{\scriptscriptstyle \rm I}=n$.

\item Multiple action of $T_{32}$
      \begin{gather}
      T_{32}(\bar w)\mathbb{B}^{a,b}(\bar u;\bar v)=\lambda_2(\bar w) \sum{\mathsf r}
      _3(\bar\xi_{\scriptscriptstyle\rm I})\,\frac{{\sf f}(\bar\xi_{\scriptscriptstyle\rm I}
      ,\bar\xi_{\scriptscriptstyle\rm I\hspace{-1pt}I}){\sf f}(\bar\xi_{\scriptscriptstyle\rm I}
      ,\bar\xi_{\scriptscriptstyle\rm I\hspace{-1pt}I\hspace{-1pt}I}){\sf f}
      (\bar\xi_{\scriptscriptstyle\rm I\hspace{-1pt}I\hspace{-1pt}I},\bar\xi_{\scriptscriptstyle\rm I\hspace{-1pt}
      I}){\sf f}(\bar\eta_{\scriptscriptstyle\rm I},\bar\eta_{\scriptscriptstyle\rm I\hspace{-1pt}I})}{{\sf f}
      (\bar\xi_{\scriptscriptstyle\rm I},\bar\eta){\sf f}(\bar\eta_{\scriptscriptstyle\rm I},\bar w){\sf f}
      (\bar w,\bar\xi_{\scriptscriptstyle\rm I\hspace{-1pt}I})}
      \nonumber
      \\
      \qquad
     {} \times{\sf K}^{(l)}_n(\bar\eta_{\scriptscriptstyle\rm I}|\bar w)\,{\sf K}^{(l)}
      _n(\bar\xi_{\scriptscriptstyle\rm I}|\bar\eta_{\scriptscriptstyle\rm I})\,{\sf K}^{(r)}
      _n(\bar w|\bar\xi_{\scriptscriptstyle\rm I\hspace{-1pt}I})\,\,\mathbb{B}^{a,b-n}
      (\bar\eta_{\scriptscriptstyle\rm I\hspace{-1pt}I};\bar\xi_{\scriptscriptstyle\rm I\hspace{-1pt}
      I\hspace{-1pt}I}).
      \label{act32}
      \end{gather}
      The sum is taken over partitions of:
      $\bar\xi\Rightarrow\{\bar\xi_{\scriptscriptstyle \rm
      I},\bar\xi_{\scriptscriptstyle \rm I\hspace{-1pt}I},\bar\xi_{\scriptscriptstyle \rm
      I\hspace{-1pt}I\hspace{-1pt}I}\}$ with $\#\bar\xi_{\scriptscriptstyle \rm I}=\#\bar\xi_{\scriptscriptstyle
      \rm I\hspace{-1pt}I}=n$;
      $\bar\eta\Rightarrow\{\bar\eta_{\scriptscriptstyle \rm I},\bar\eta_{\scriptscriptstyle \rm
      I\hspace{-1pt}I}\}$ with $\#\bar\eta_{\scriptscriptstyle \rm I}=n$.
\item Multiple action of $T_{31}$
      \begin{gather}
      T_{31}(\bar w)\mathbb{B}^{a,b}(\bar u;\bar v)=\lambda_2(\bar w) \sum{\mathsf r}
      _1(\bar\eta_{\scriptscriptstyle\rm I\hspace{-1pt}I})\,{\mathsf r}_3(\bar\xi_{\scriptscriptstyle\rm I}
      )\,{\sf K}^{(l)}_n(\bar\xi_{\scriptscriptstyle\rm I}|\bar\eta_{\scriptscriptstyle\rm I})\,{\sf K}^{(r)}
      _n(\bar\xi_{\scriptscriptstyle\rm I\hspace{-1pt}I}|\bar\eta_{\scriptscriptstyle\rm I\hspace{-1pt}I}
      )\,{\sf K}^{(l)}_n(\bar\eta_{\scriptscriptstyle\rm I}|\bar w)\,{\sf K}^{(r)}
      _n(\bar w|\bar\xi_{\scriptscriptstyle\rm I\hspace{-1pt}I})
      \nonumber
      \\
      \qquad
      {}\times\frac{{\sf f}(\bar\eta_{\scriptscriptstyle\rm I},\bar\eta_{\scriptscriptstyle\rm I\hspace{-1pt}I}
      ){\sf f}(\bar\eta_{\scriptscriptstyle\rm I},\bar\eta_{\scriptscriptstyle\rm I\hspace{-1pt}I\hspace{-1pt}I}
      ){\sf f}(\bar\eta_{\scriptscriptstyle\rm I\hspace{-1pt}I\hspace{-1pt}I}
      ,\bar\eta_{\scriptscriptstyle\rm I\hspace{-1pt}I}){\sf f}(\bar\xi_{\scriptscriptstyle\rm I}
      ,\bar\xi_{\scriptscriptstyle\rm I\hspace{-1pt}I}){\sf f}(\bar\xi_{\scriptscriptstyle\rm I}
      ,\bar\xi_{\scriptscriptstyle\rm I\hspace{-1pt}I\hspace{-1pt}I}){\sf f}
      (\bar\xi_{\scriptscriptstyle\rm I\hspace{-1pt}I\hspace{-1pt}I},\bar\xi_{\scriptscriptstyle\rm I\hspace{-1pt}
      I})}{{\sf f}(\bar\xi_{\scriptscriptstyle\rm I},\bar\eta){\sf f}
      (\bar\xi_{\scriptscriptstyle\rm I\hspace{-1pt}I\hspace{-1pt}I}
      ,\bar\eta_{\scriptscriptstyle\rm I\hspace{-1pt}I}){\sf f}(\bar\xi_{\scriptscriptstyle\rm I\hspace{-1pt}I}
      ,\bar\eta_{\scriptscriptstyle\rm I\hspace{-1pt}I}){\sf f}(\bar\eta_{\scriptscriptstyle\rm I},\bar w){\sf f}
      (\bar w,\bar\xi_{\scriptscriptstyle\rm I\hspace{-1pt}I})}\,\mathbb{B}^{a-n,b-n}
      (\bar\eta_{\scriptscriptstyle\rm I\hspace{-1pt}I\hspace{-1pt}I}
      ;\bar\xi_{\scriptscriptstyle\rm I\hspace{-1pt}I\hspace{-1pt}I}).
      \label{act31}
      \end{gather}
      The sum is taken over partitions of:
      $\bar\xi\Rightarrow\{\bar\xi_{\scriptscriptstyle \rm
      I},\bar\xi_{\scriptscriptstyle \rm I\hspace{-1pt}I},\bar\xi_{\scriptscriptstyle \rm
      I\hspace{-1pt}I\hspace{-1pt}I}\}$ with $\#\bar\xi_{\scriptscriptstyle \rm I}=\#\bar\xi_{\scriptscriptstyle
      \rm I\hspace{-1pt}I}=n$;
      $\bar\eta\Rightarrow\{\bar\eta_{\scriptscriptstyle \rm I},\bar\eta_{\scriptscriptstyle \rm
      I\hspace{-1pt}I},\bar\eta_{\scriptscriptstyle \rm I\hspace{-1pt}I\hspace{-1pt}I}\}$ with
      $\#\bar\eta_{\scriptscriptstyle \rm I}=\#\bar\eta_{\scriptscriptstyle \rm I\hspace{-1pt}I}=n$.
\end{itemize}
\end{prop}

Note that the product of the rational functions ${\sf f}(\bar\xi_{\scriptscriptstyle \rm I},\bar\eta) {\sf
f}(\bar\xi_{\scriptscriptstyle \rm I\hspace{-1pt}I\hspace{-1pt}I},\bar\eta_{\scriptscriptstyle \rm
I\hspace{-1pt}I}) {\sf f}(\bar\xi_{\scriptscriptstyle \rm I\hspace{-1pt}I},\bar\eta_{\scriptscriptstyle \rm
I\hspace{-1pt}I})$ in the denominator of the r.h.s.\
of~\eqref{act31} can be equally rewritten as ${\sf f}(\bar\xi,\bar\eta_{\scriptscriptstyle \rm
I\hspace{-1pt}I}) {\sf f}(\bar\xi_{\scriptscriptstyle \rm I},\bar\eta_{\scriptscriptstyle \rm I}) {\sf
f}(\bar\xi_{\scriptscriptstyle \rm I},\bar\eta_{\scriptscriptstyle \rm I\hspace{-1pt}I\hspace{-1pt}I})$.

The proof of formulae~\eqref{act13}--\eqref{act31} will be divided into two steps.
First, we will prove these formulae using the current approach and presentation of the of\/f-shell Bethe
vectors in the form~\eqref{rbv2} for the action of only one monodromy element, that is $\#\bar w=n=1$.
Then we will use an induction to prove these formulae for $n>1$.

\section{Proofs}

In what follows we will identify the monodromy matrix $T(u)$ with the $L$-operator
${L}^+(u)\in U_q(\mathfrak{b}^+)$ from the positive Borel subalgebra of the quantum af\/f\/ine algebra
$\Uq{3}$.

\subsection[The case $\#\bar w=1$]{The case $\boldsymbol{\#\bar w=1}$}

As we have already mentioned our f\/irst goal is the proof of the action
formulae~\eqref{act13}--\eqref{act31} for the single action of the monodromy matrix elements onto
of\/f-shell Bethe vectors.
In this subsection, we perform this calculation using only the commutation relations of $\Uq{3}$ current
generators.

\subsubsection{Necessary commutation relations}

Since the essential part of the of\/f-shell Bethe vectors is concentrated in the projection of full current
products, we may consider f\/irst the action of monodromy elements onto the projection of a~special product
of the full currents.

According to the properties of the projections~\eqref{hF-ord-a} we can present the projection
${P}^+_f({\cal F}_2(\bar v)$ ${\cal F}_1(\bar u))$ in the form
\begin{gather}
\label{F-int}
{P}^+_f\sk{{\cal F}_2(\bar v){\cal F}_1(\bar u)}={\cal F}_2(\bar v){\cal F}_1(\bar u)-\sum{P}
^-_f\sk{{\cal F}''}\cdot{P}^+_f\sk{{\cal F}'},
\end{gather}
where the elements ${\cal F}'$ and ${\cal F}''$ are def\/ined by the coproduct~\eqref{gln-cm}
\begin{gather*}
\Delta^{{(D)}}\sk{{\cal F}_2(\bar v){\cal F}_1(\bar u)}=\sum{\cal F}'\otimes{\cal F}'',
\end{gather*}
and in the r.h.s.\
of~\eqref{F-int} the number of currents entering the elements ${\cal F}'$ is less than the total number of
currents in the original product ${\cal F}_2(\bar v){\cal F}_1(\bar u)$.
Then we may continue replacing ${P}^+_f\sk{{\cal F}'}$ by the r.h.s.\
of~\eqref{F-int} up to the trivial identity ${P}^+_f\sk{{F}_i(w)}={F}_i(w)-{P}^-_f\sk{{F}_i(w)}$ to obtain
the presentation of ${P}^+_f\sk{{\cal F}_2(\bar v){\cal F}_1(\bar u)}$ as a~linear combination of terms
which are ordered products of negative projections of the currents and the full currents.
The idea of calculation of the action of the monodromy elements is to act on this sum f\/irst and then
apply the projection ${P}^+_f$ to the result.
It will be shown below that a~lot of terms in this sum disappear.
Then, it is easy to control the surviving terms.

Let $I$ be the right ideal of $\Uq{3}$ generated by all elements of the form ${F}_i[n]\cdot \Uq{3}$ for
$i=1,2$ and $n<0$.
We will denote equalities modulo elements in the ideal $I$ by the symbol `$\sim_I$'.
Note that this ideal is annihilated by the projection ${P}^+_f$.

A useful presentation of the of\/f-shell Bethe vector was proved in the paper~\cite{FKPR} using the notion
of $q$-deformed symmetrization (see Corollary 3.6 in that paper).
We rewrite this presentation replacing deformed symmetrization by usual symmetrization (with multiplication
by a~scalar factor).
We have\footnote{The reasons for existence of the presentation~\eqref{us-P2} were explained in the
paper~\cite{KP-GS}, where the whole inf\/inite set of the hierarchical relations between $\Uq{N}$
of\/f-shell Bethe vectors was described in terms of the generating series.}~\cite{FKPR,KP-GS}
\begin{gather}
{P}^+_f({\cal F}_2(\bar v){\cal F}_1(\bar u))
={\cal F}_2(\bar v)\cdot{\cal F}_1(\bar u)
-\sum_{i=1}^b{P}^-_f\skk{{F}_{3,2}(v_i)}\cdot{\cal F}_2(\bar v_i)\cdot{\cal F}_1(\bar u)
\frac{{\sf f}(v_i,\bar v_{>i})}{{\sf f}(\bar v_{>i},v_i)}
\nonumber
\\
\qquad{}
-\sum_{i=1}^a{P}^-_f\skk{{F}_{2,1}(u_i)}\cdot{\cal F}_2(\bar v)\cdot{\cal F}_1(\bar u_i){\sf f}
(\bar v,u_i)\frac{{\sf f}(u_i,\bar u_{>i})}{{\sf f}(\bar u_{>i},u_i)}
\label{us-P2}
\\
\qquad{}
- \sum_{\substack{1\leq i\leq b\\1\leq j\leq a}}
\frac{{P}^-_f\skk{{F}_{3,1}(u_j)}}{q-q^{-1}}
\cdot{\cal F}_2(\bar v_i)\cdot{\cal F}_1(\bar u_j)
{\sf g}(v_i,u_j)v_i
{\sf f}(\bar v_i,u_j)
\frac{{\sf f}(v_i,\bar v_{>i})}{{\sf f}
(\bar v_{>i},v_i)}\frac{{\sf f}(u_j,\bar u_{>j})}{{\sf f}(\bar u_{>j},u_j)}+\mathbb{W}, \nonumber
\end{gather}
where the elements $\mathbb{W}$ are such that ${P}^+_f\sk{T_{ij}(w)\cdot \mathbb{W}}=0$.
Recall that $\bar v_i$ and $\bar u_j$ are the sets $\bar v\setminus\{v_i\}$ and $\bar u\setminus\{u_j\}$.
This fact will be checked further using an equivalence
\begin{gather}
\label{ac-m-fm}
T_{ij}(w)\cdot{P}^-_f\skk{{F}_{k,l}(u)}\sim_I\delta_{i,k}\big(q-q^{-1}\big)^{k-l-1}{\sf g}(w,u)\,uT_{lj}(w),
\end{gather}
also proved in~\cite{FKPR}.
Here and in~\eqref{us-P2} the notation ${F}_{k,l}(u)$, $1\leq l<k\leq 3$ is used to denote the simple and
`composed' currents (see~\eqref{com-cur} and discussion on the `analytical' properties of the composed
currents in~\cite{FKPR, KP-GLN}):
\begin{gather*}
{F}_{2,1}(u)\equiv F_1(u),
\qquad
{F}_{3,2}(u)\equiv F_2(u),
\qquad
{F}_{3,1}(u)\equiv\big(q-q^{-1}\big)F_1(u)F_2(u).
\end{gather*}

The equivalence~\eqref{ac-m-fm} allows one to prove easily that ${P}^+_f\sk{T_{ij}(w)\cdot \mathbb{W}}=0$
since the elements of $\mathbb{W}$ can be presented in general as $\sum {P}^-_f\sk{{F}_{c_1,k}}\cdot
{P}^-_f\sk{{F}_{c_2,l}}\cdot \mathbb{W}'$ with $c_1>k$ and $c_2>l$.
For example, for $k=l=1$ and according to~\eqref{ac-m-fm} the action $T_{ij} \cdot
{P}^-_f\sk{{F}_{c_1,1}}\cdot {P}^-_f\sk{{F}_{c_2,1}}\cdot \mathbb{W}'$ is proportional to
$\delta_{i,c_1}\delta_{1,c_2}=0$ since $c_2>1$.
This means that the action of the elements of the monodromy elements onto universal of\/f-shell Bethe
vectors is def\/ined only by the four terms presented in~\eqref{us-P2}.
Then, the calculation of this action will be reduced to the commutation of Gauss coordinates entering the
monodromy elements~\eqref{Gaudec} and the full currents, which is relatively simple.

The calculation of the action of the monodromy matrix elements onto the Bethe vector ${P}^+_f({\cal
F}_2(\bar v){\cal F}_1(\bar u))$ is decomposed in several steps.
First we use formula~\eqref{ac-m-fm} to get rid of the negative projection of the currents and obtain
products of the monodromy elements and the full currents.
Then we use the explicit expressions of the monodromy matrix elements~\mbox{\eqref{GF2}--\eqref{GE2}} through the
Gauss coordinates to calculate the commutation of the Gauss coordina\-tes~${\rm E}^+_{ij}(w)$, ${k}^+_i(w)$
and the full currents, calculating this commutation modulo certain ideals $J$ and $K$ which will be
described below.
In the next step, we apply the projection ${P}^+_f$ to the result of this calculation to restore the
structure of the of\/f-shell Bethe vectors, using formula~\eqref{rbv2}.
Finally, we rewrite the resulting sum of Bethe vectors as a~sum over partitions.

To proceed further, we need to know the commutation relations between the Gauss coordinates ${\rm
E}^+_{ij}(w)$ and the full currents $F_i(u)$.
To identify ${P}^+_f({\cal F}_2(\bar v){\cal F}_1(\bar u))$ with the of\/f-shell Bethe vector we have to
act with this element on the right weight singular vector $|0\rangle$.
Thus, we can perform the calculations modulo the right ideal $J$ composed from elements
$U_q(\widehat{\mathfrak{gl}}_3)\cdot {E}_i[n]$ for $i=1,2$ and $n\geq 0$.
Moreover, the commutation relations of ${\rm E}^+_{ij}(u)$ with the full currents ${F}_i(u)$ produce terms
containing the negative Cartan currents ${k}^-(u)$ which can be neglected since they vanish after
application of the projection ${P}^+_f$.
We note $K$ the ideal formed by such elements and equalities
modulo elements of the ideals $J$ and $K$ will
be denoted by `$\sim_J$' and `$\sim_K$' respectively.                     

In what follows we need to express the Gauss coordinate ${\rm E}^+_{13}(w)$ through the current generators.
From the $RLL$-relation~\eqref{YB} one can obtain the relation
\begin{gather}
\label{YBadd}
(v-u)[{L}^-_{21}(u),{L}^+_{32}(v)]=\big(q-q^{-1}\big)\sk{u{L}^+_{22}(v){L}^-_{31}(u)-v{L}^-_{22}(u){L}^+_{31}(v)}.
\end{gather}
According to the def\/inition~\eqref{Lop}, ${L}^-_{ij}(u)$ are series with respect to non-negative powers
of the spectral parameter $u$.
The coef\/f\/icient at $u^0$ in~\eqref{YBadd} yields the following relation
\begin{gather}
\label{YBa1}
\big(q-q^{-1}\big){L}^-_{22}[0]{L}^+_{31}(v)=-[{L}^-_{21}[0],{L}^+_{32}(v)].
\end{gather}
Next we use the explicit expression of the $L$-operator matrix elements in terms of the Gauss
coordinates~\eqref{GE2} and the inverted Ding--Frenkel formulae~\eqref{DFinverse} to observe that
\begin{gather}
\label{obser}
{L}^-_{21}[0]=-k^-_2[0]E_1[0],
\qquad
{L}^-_{22}[0]=k^-_2[0],
\qquad
{L}^+_{3i}(v)=k^+_3(v){\rm E}^+_{i3}(v),
\quad
i=1,2.
\end{gather}
Let us remind that by def\/inition the Gauss coordinate ${\rm E}^+_{23}(w)$ coincides with the projection
of the simple root currents $E_2(w)$ (see~\eqref{DFinverse})
\begin{gather}
\label{EEsr}
{\rm E}^+_{23}(v)={P}^+_e\sk{E_2(v)}=\sum_{n>0}E_2[n]v^{-n}=\oint\frac{dt}{v}\frac{E_2(t)}{1-t/v},
\qquad
i=1,2.
\end{gather}
Substituting
the relations~\eqref{obser} into~\eqref{YBa1} and using the commutation relations
$E_{2}(t)k^-_2[0]=qk^-_2[0]E_{2}(t)$ and $k^+_3(v)E_1[0]=E_1[0]k^+_3(v)$ that follow from~\eqref{kiE}
and~\eqref{kE} respectively we obtain f\/inally
\begin{gather}
\label{E13int}
{\rm E}^+_{13}(w)=\frac{1}{q-q^{-1}}\oint\frac{dt}{w(1-t/w)}\sk{E_{1}[0]E_{2}(t)-qE_{2}(t)E_{1}[0]}.
\end{gather}
In~\eqref{EEsr} and~\eqref{E13int} the symbol $\oint dt\, g(t)$ means the term $g_{-1}$ of the formal
series $g(t)=\sum\limits_{n\in{\mathbb Z}}g_nt^{-n}$ and the rational function $\frac{1}{1-t/v}$ is
understood as a~series $\sum\limits_{n\geq 0}(t/v)^n$.

Then from~\eqref{EF} we observe that
\begin{gather*}
[{\rm E}^+_{i\,i+1}(w),{F}_j(u)]\sim_K\delta_{ij}{\sf g}(w,u)u\psi^+_i(u),
\qquad
[E_i[0],{F}_j(u)]\sim_K\delta_{ij}\big(q-q^{-1}\big)\psi^+_i(u),
\\
\psi^+_i(u)={k}^+_i(u)/{k}^+_{i+1}(u),
\qquad
i=1,2.
\end{gather*}
Using also one more relation
\begin{gather*}
E_{1}[0]\psi^+_{2}(w)-q\psi^+_{2}(w)E_{1}[0]=\big(q-q^{-1}\big)\psi^+_2(w){\rm E}^+_{12}(w),
\end{gather*}
which follows from~\eqref{ki1E} and~\eqref{kE}, we may conclude that the action of the Gauss coordina\-tes~${\rm E}^+_{ij}(u)$ onto the product of the full currents ${\cal F}_2(\bar v){\cal F}_1(\bar u)$ is given
by the equalities
\begin{gather}
{\rm E}^+_{13}(w)\cdot{\cal F}_2(\bar v){\cal F}_1(\bar u)\sim_{K,J}
\sum_{\substack{1\leq i\leq b\\1\leq j\leq a}}
{\cal F}_2(\bar v_i){\cal F}_1(\bar u_j)\psi^+_2(v_i)\psi^+_1(u_j)
\nonumber
\\
\hphantom{{\rm E}^+_{13}(w)\cdot{\cal F}_2(\bar v){\cal F}_1(\bar u)\sim_{K,J}}{}
\times{\sf g}(w,v_i)v_i{\sf g}(v_i,u_j)u_j{\sf f}(v_i,\bar u_j)\frac{{\sf f}(\bar u_{<j},u_j)}{{\sf f}
(u_j,\bar u_{<j})}\frac{{\sf f}(\bar v_{<i},v_i)}{{\sf f}(v_i,\bar v_{<i})},
\label{ae13-1}
\\
\label{ae12-1}
{\rm E}^+_{12}(w)\cdot{\cal F}_2(\bar v){\cal F}_1(\bar u)\sim_{K,J}\sum_{j=1}^a{\cal F}_2(\bar v){\cal F}
_1(\bar u_j)\psi^+_1(u_j){\sf g}(w,u_j)u_j
\frac{{\sf f}(\bar u_{<j},u_j)}{{\sf f}(u_j,\bar u_{<j})},
\\
\label{ae23-1}
{\rm E}^+_{23}(w)\cdot{\cal F}_2(\bar v){\cal F}_1(\bar u)\sim_{K,J}\sum_{i=1}^b{\cal F}
_2(\bar v_i){\cal F}_1(\bar u)\psi^+_2(v_i){\sf g}(w,v_i)v_i{\sf f}(v_i,\bar u)\frac{{\sf f}(\bar v_{<i}
,v_i)}{{\sf f}(v_i,\bar v_{<i})}.
\end{gather}

Now that we have established the action of the Gauss coordinates on products of the full current, we can
compute the action of the monodromy operators on Bethe vectors.

\subsubsection{Calculation of the action}

$\bullet$ The action of $T_{13}(w)$.
Let us specialize the vector $\mathbb{B}^{a+1,b+1}(w,\bar u;\bar v,w')$ given by the
expression~\eqref{rbv2} at the coinciding points $w'=w$.
We have
\begin{gather}
{\mathbb{B}}^{a+1,b+1}(w,\bar u;\bar v,w')|_{w'=w}=\frac{\beta(\bar u|\bar v)}{{\sf f}(\bar v,\bar u)}
\frac{{\sf f}(\bar v,w'){\sf f}(w,\bar u)}{{\sf f}(\bar v,w){\sf f}(w',\bar u)}{\sf r}_3(\bar v){\sf r}
_3(w')
\nonumber
\\
\left.\qquad{}
\times
\frac{w'-w}{qw'-q^{-1}w}{P}^+_f\left({F}_{2}(v_b)\cdots{F}_{2}(v_1){F}_2(w')\cdot{F}_1(w){F}_{1}
(u_a)\cdots{F}_{1}(u_1)\right)\right|_{w'=w}|0\rangle.
\label{c13}
\end{gather}
Using the commutation relations~\eqref{FiFi1}, the r.h.s.\
of~\eqref{c13} can be written as
\begin{gather}
{\mathbb{B}}^{a+1,b+1}(w,\bar u;\bar v,w)=\frac{\beta(\bar u|\bar v)}{{\sf f}(\bar v,\bar u)}{\sf r}
_3(\bar v){\sf r}_3(w)
\nonumber
\\
\hphantom{{\mathbb{B}}^{a+1,b+1}(w,\bar u;\bar v,w)=}{}
\times{P}^+_f\left({F}_{2}(v_b)\cdots{F}_{2}(v_1){F}_1(w)\cdot{F}_2(w){F}_{1}(u_a)\cdots{F}_{1}
(u_1)\right)|0\rangle.
\label{c13-1}
\end{gather}

On the other hand, the action of the elements $T_{13}(w)$, according to the property~\eqref{ac-m-fm}, is
given only by the f\/irst term in the r.h.s.\
of~\eqref{us-P2}, namely by the product of the full currents ${\cal F}_2(\bar v)\cdot{\cal F}_1(\bar u)$,
so that using the explicit form $T_{13}(w)={\rm F}^+_{31}(w)k^+_3(w)$ we can write
\begin{gather}
T_{13}(w){\mathbb{B}}^{a,b}(\bar u;\bar v)=\frac{\beta(\bar u|\bar v)}{{\sf f}(\bar v,\bar u)}{\sf r}_3(\bar v)
\nonumber
\\
\hphantom{T_{13}(w){\mathbb{B}}^{a,b}(\bar u;\bar v)=}{}
\times {P}^+_f\left({\rm F}^+_{31}(w)k^+_3(w){F}_{2}(v_b)\cdots{F}_{2}(v_1)\cdot{F}_{1}(u_a)\cdots{F}_{1}
(u_1)\right)|0\rangle.\label{c13-2}
\end{gather}
Taking into account the relation between the Gauss coordinate ${\rm F}^+_{31}(w)$ and the projection of the
composed current ${F}_{3,1}(w)=\big(q-q^{-1}\big){F}_1(w){F}_2(w)$~\cite{KP-GLN}
\begin{gather*}
{P}^+_f\sk{{F}_{3,1}(w)}=\big(q-q^{-1}\big){\rm F}^+_{31}(w)
\qquad
\text{or}
\qquad
{\rm F}^+_{31}(w)={P}^+_f\sk{{F}_1(w){F}_2(w)},
\end{gather*}
the property of the projection operator
\begin{gather}
\label{pro-pro}
{P}^+_f\sk{{P}^+_f\sk{A}\cdot B}={P}^+_f\sk{A\cdot B},
\end{gather}
and the commutation relation
\begin{gather*}
{F}_1(w){F}_2(w)k_3(w)\cdot{F}_2(v)={F}_2(v)\cdot{F}_1(w){F}_2(w)k_3(w),
\end{gather*}
we conclude that the r.h.s.\
of~\eqref{c13-2} is equal to the r.h.s.\
of~\eqref{c13-1} up to multiplication by $\lambda_2(w)$ and hence the relation~\eqref{act13} is proved for
$n=1$.

$\bullet$ The action of $T_{12}(w)$.
Again, due to~\eqref{ac-m-fm}, the action of the monodromy matrix element $T_{12}(w)$ onto the Bethe
vector~\eqref{rbv2} is determined by the product of the full currents ${\cal F}_2(\bar v)\cdot{\cal
F}_1(\bar u)$.
Taking into account that
\begin{gather*}
T_{12}(w)={\rm F}^+_{21}(w){k}^+_2(w)+{\rm F}^+_{31}(w){k}^+_3(w){\rm E}^+_{23}(w)={\rm F}^+_{21}(w){k}
^+_2(w)+T_{13}(w){E}^+_{23}(w),
\end{gather*}
using~\eqref{ae23-1} and the commutation relations of the Cartan currents ${k}^+_2(w)$ with the full
currents given by~\eqref{kiF} and~\eqref{ki1F} we obtain
\begin{gather}
T_{12}(w)\mathbb{B}^{a,b}(\bar u;\bar v)=\lambda_2(w){\sf f}(\bar v,w)\mathbb{B}^{a+1,b}
(w,\{\bar u;\bar v\})
\nonumber
\\
\hphantom{T_{12}(w)\mathbb{B}^{a,b}(\bar u,\bar v)=}
{} +T_{13}(w)\sum_{i=1}^b{\sf K}^{(r)}_1(w|v_i){\sf f}(\bar v_i,v_i)\mathbb{B}^{a,b-1}(\bar u;\bar v_i).
\label{c12-1}
\end{gather}
In~\eqref{c12-1} we replace the function ${\sf g}(w,v_i)v_i$ by the function ${\sf K}^{(r)}_1(w|v_i)$
using~\eqref{Mod-Izer} and~\eqref{21}.
In the f\/irst term of the r.h.s.\
of~\eqref{c12-1} we used again the property of the projection~\eqref{pro-pro} and the commutation relation
\begin{gather*}
{F}_1(w)k^+_2(w)\cdot F_2(v)=F_2(v)\cdot{F}_1(w)k^+_2(w).
\end{gather*}
Using the action of $T_{13}(w)$ onto the of\/f-shell Bethe vector (just calculated above) we may
rewrite~\eqref{c12-1} in the form
\begin{gather}
T_{12}(w)\mathbb{B}^{a,b}(\bar u;\bar v)=\lambda_2(w){\sf f}(\bar v,w)\mathbb{B}^{a+1,b}(\{w,\bar u\}
;\bar v)
\nonumber
\\
\hphantom{T_{12}(w)\mathbb{B}^{a,b}(\bar u,\bar v)=}
{} +\lambda_2(w)\sum_{i=1}^b{\sf K}^{(r)}_1(w|v_i){\sf f}(\bar v_i,v_i)\mathbb{B}^{a+1,b}(\{w,\bar u\}
;\{w,\bar v_i\}),
\label{c12-2}
\end{gather}
which can be rewritten in the form~\eqref{act12} as a~sum over partitions of the set $\xi=\{w,\bar
v\}\Rightarrow \{\bar\xi_{\scriptscriptstyle \rm I},\bar\xi_{\scriptscriptstyle \rm I\hspace{-1pt}I}\}$,
for $\#\bar\xi_{\scriptscriptstyle \rm I}=1$ since
\begin{gather}
\label{reduc}
\left.\frac{{\sf K}^{(l,r)}_1(w|\bar\xi_{\scriptscriptstyle\rm I})}{{\sf f}
(w,\bar\xi_{\scriptscriptstyle\rm I})}\right|_{\bar\xi_{\scriptscriptstyle\rm I}=\{w\}}=1.
\end{gather}
The action~\eqref{act12} for $n=1$ is proved.

$\bullet$  The action of $T_{23}(w)$.
According to~\eqref{ac-m-fm} the action of the monodromy matrix element $T_{23}(w)={\rm
F}^+_{32}(w)k^+_3(w)$ will be def\/ined by the f\/irst and third terms of the r.h.s.\
of~\eqref{us-P2} which produce two terms in the action:
\begin{gather*}
T_{23}(w)\mathbb{B}^{a,b}(\bar u;\bar v)=\lambda_2(w){\sf f}(w,\bar u)\mathbb{B}^{a,b+1}
(\bar u;\{\bar v,w\})
\\
\hphantom{T_{23}(w)\mathbb{B}^{a,b}(\bar u,\bar v)=}
{} -T_{13}(w)\sum_{j=1}^a{\sf g}(w,u_j)u_j{\sf f}(u_j,\bar u_j)\mathbb{B}^{a-1,b}(\bar u_j;\bar v),
\end{gather*}
or
\begin{gather*}
T_{23}(w)\mathbb{B}^{a,b}(\bar u;\bar v)=\lambda_2(w){\sf f}(w,\bar u)\mathbb{B}^{a,b+1}
(\bar u;\{\bar v,w\})
\\
\hphantom{T_{23}(w)\mathbb{B}^{a,b}(\bar u,\bar v)=}
{} +\lambda_2(w)\sum_{j=1}^a{\sf K}^{(l)}_1(u_j|w){\sf f}(u_j,\bar u_j)\mathbb{B}^{a,b+1}(\{\bar u_j,w\}
;\{\bar v,w\}).
\end{gather*}
Due to~\eqref{reduc}, they can be rewritten as the sum over partition of the set $\eta=\{w,\bar
u\}\Rightarrow \{\bar\eta_{\scriptscriptstyle \rm I},\bar\eta_{\scriptscriptstyle \rm I\hspace{-1pt}I}\}$,
for $\#\bar\eta_{\scriptscriptstyle \rm I}=1$.
The action~\eqref{act23} for $n=1$ is proved.

$\bullet$ The action of $T_{22}(w)$.
The action of the matrix element
\begin{gather*}
T_{22}(w)={k}^+_2(w)+{\rm F}^+_{32}(w){k}^+_3(w){\rm E}^+_{23}(w)
\end{gather*}
onto the of\/f-shell Bethe vector~\eqref{rbv2} is determined according to~\eqref{ac-m-fm} by the f\/irst
and the third terms in~\eqref{us-P2} and using~\eqref{ae23-1} we obtain
\begin{gather}
T_{22}(w)\mathbb{B}^{a,b}(\bar u;\bar v)=\lambda_2(w){\sf f}(w,\bar u){\sf f}(\bar v,w)\mathbb{B}^{a,b}
(\bar u;\bar v)
\nonumber
\\
\hphantom{T_{22}(w)\mathbb{B}^{a,b}(\bar u,\bar v)=}{}
+\lambda_2(w){\sf f}(w,\bar u)\sum_{i=1}^b{\sf g}(w,v_i)v_i{\sf f}(\bar v_i,v_i)\mathbb{B}^{a,b}
(\bar u;\{\bar v_i,w\})
\nonumber
\\
\hphantom{T_{22}(w)\mathbb{B}^{a,b}(\bar u,\bar v)=}{}
+\sum_{j=1}^a{\sf g}(u_j,w)u_j{\sf f}(u_j,\bar u_j)T_{12}(w)\mathbb{B}^{a-1,b}(\bar u_j;\bar v).
\label{c22}
\end{gather}
Using now the explicit formula~\eqref{c12-2} for the action of the monodromy matrix element $T_{12}(w)$
onto the of\/f-shell Bethe vector we may rewrite~\eqref{c22} in the form
\begin{gather}
T_{22}(w)\mathbb{B}^{a,b}(\bar u;\bar v)=\lambda_2(w){\sf f}(w,\bar u){\sf f}(\bar v,w)\mathbb{B}^{a,b}
(\bar u;\bar v)
\nonumber
\\
\qquad
{}+\lambda_2(w){\sf f}(w,\bar u)\sum_{i=1}^b{\sf K}^{(r)}_1(w|v_i){\sf f}(\bar v_i,v_i)\mathbb{B}^{a,b}
(\bar u;\{\bar v_i,w\})
\nonumber
\\
\qquad
{}+\lambda_2(w){\sf f}(\bar v,w)\sum_{j=1}^a{\sf K}^{(l)}_1(u_j|w){\sf f}(u_j,\bar u_j)\mathbb{B}^{a,b}
(\{\bar u_j,w\};\bar v)
\nonumber
\\
\qquad
{}+\lambda_2(w)\sum_{\substack{1\leq i\leq b\\1\leq j\leq a}}
{\sf K}^{(l)}_1(u_j|w){\sf K}^{(r)}_1(w|v_i){\sf f}(\bar v_i,v_i){\sf f}
(u_j,\bar u_j)\mathbb{B}^{a,b}(\{\bar u_j,w\};\{\bar v_i,w\}),
\label{c22-1}
\end{gather}
which can be presented as sum over partitions~\eqref{act22} of the sets
\begin{gather}
\label{part1}
\eta=\{w,\bar u\}\Rightarrow\{\bar\eta_{\scriptscriptstyle\rm I},\bar\eta_{\scriptscriptstyle\rm I\hspace{-1pt}I}\}
\qquad
\text{and}
\qquad
\xi=\{w,\bar v\}\Rightarrow\{\bar\xi_{\scriptscriptstyle\rm I},\bar\xi_{\scriptscriptstyle\rm I\hspace{-1pt} I}\}
\qquad
\text{for}  
\qquad
\#\bar\eta_{\scriptscriptstyle\rm I}=\#\bar\xi_{\scriptscriptstyle\rm I}=1.
\end{gather}
The action~\eqref{act22} for $n=1$ is proved.

$\bullet$ The action of $T_{11}(w)$.
The action of the matrix element
\begin{gather*}
T_{11}(w)={k}^+_1(w)+{\rm F}^+_{21}(w){k}^+_2(w){\rm E}^+_{12}(w)+{\rm F}^+_{31}(w){k}^+_3(w){\rm E}^+_{13}
(w)
\\
\hphantom{T_{11}(w)}{}
={k}^+_1(w)+{\rm F}^+_{21}(w){k}^+_2(w){\rm E}^+_{12}(w)+T_{13}(w){E}^+_{13}(w)
\end{gather*}
as well as the matrix elements $T_{12}(w)$ and $T_{13}(w)$ is determined due to~\eqref{ac-m-fm} by the
f\/irst term in~\eqref{us-P2}.
Using formulae~\eqref{ae13-1} and~\eqref{ae12-1} we obtain
\begin{gather}
T_{11}(w)\mathbb{B}^{a,b}(\bar u;\bar v)=\lambda_2(w){\sf r}_1(w){\sf f}(\bar u,w)\mathbb{B}^{a,b}
(\bar u;\bar v)
\nonumber
\\
\qquad{}
+\lambda_2(w)\sum_{j=1}^a{\sf r}_1(u_j){\sf K}^{(r)}_1(w|u_j)\frac{{\sf f}(\bar u_j,u_j){\sf f}
(\bar v,w)}{{\sf f}(\bar v,u_j)}\mathbb{B}^{a,b}(\{\bar u_j,w\};\bar v)
\nonumber
\\
\qquad{}
+\lambda_2(w)\sum_{\substack{1\leq i\leq b\\1\leq j\leq a}}
{\sf r}_1(u_j){\sf K}^{(r)}_1(w|v_i){\sf K}^{(r)}_1(v_i|u_j)\frac{{\sf f}
(\bar u_j,u_j){\sf f}(\bar v_i,v_i)}{{\sf f}(\bar v,u_j)}\mathbb{B}^{a,b}(\{\bar u_j,w\};\{\bar v_i,w\}).\!\!\!
\label{c11}
\end{gather}
The expression~\eqref{c11} can be written as the sum~\eqref{act11} over partitions~\eqref{part1}, because
the term corresponding to the partition $\xi_{\scriptscriptstyle \rm I\hspace{-1pt}I}=\{\bar v_i,w\}$ and
$\eta_{\scriptscriptstyle \rm I}=\{w\}$ vanishes due to presence of the factor ${\sf
f}(\xi_{\scriptscriptstyle \rm I\hspace{-1pt}I},\eta_{\scriptscriptstyle \rm I})$ in the denominator
of~\eqref{act11}.
The other three types of partitions $\xi_{\scriptscriptstyle \rm I}=\eta_{\scriptscriptstyle \rm I}=\{w\}$;
$\xi_{\scriptscriptstyle \rm I}=\{w\}$, $\eta_{\scriptscriptstyle \rm I}=\{u_j\}$; $\xi_{\scriptscriptstyle
\rm I}=\{v_i\}$, $\eta_{\scriptscriptstyle \rm I}=\{u_j\}$ yield exactly the three terms in~\eqref{c11} due
to~\eqref{reduc}.
The action~\eqref{act11} for $n=1$ is proved.

$\bullet$ The action of $T_{33}(w)$.
According to~\eqref{ac-m-fm} this action will be determined by the f\/irst, the second and the forth terms
in~\eqref{us-P2}.
Using these relations, the def\/inition of the universal of\/f-shell Bethe vector~\eqref{rbv2} and the fact
that $T_{33}(w)=k^+_3(w)$ we obtain
\begin{gather}
T_{33}(w)\mathbb{B}^{a,b}(\bar u;\bar v)=\lambda_2(w){\sf r}_3(w){\sf f}(w,\bar v)\mathbb{B}^{a,b}
(\bar u;\bar v)
\nonumber
\\
\qquad{}
+\lambda_2(w)\sum_{i=1}^b{\sf r}_3(v_i){\sf K}^{(l)}_1(v_i|w)\frac{{\sf f}(v_i,\bar v_i){\sf f}
(w,\bar u)}{{\sf f}(v_i,\bar u)}\mathbb{B}^{a,b}(\bar u;\{\bar v_i,w\})
\nonumber
\\
\qquad{}
+\lambda_2(w)\sum_{\substack{1\leq i\leq b\\1\leq j\leq a}}
{\sf r}_3(v_i){\sf K}^{(l)}_1(u_j|w){\sf K}^{(l)}_1(v_i|u_j)\frac{{\sf f}
(v_i,\bar v_i){\sf f}(u_j,\bar u_j)}{{\sf f}(v_i,\bar u)}\mathbb{B}^{a,b}(\{\bar u_j,w\};\{\bar v_i,w\}).\!\!\!
\label{c33}
\end{gather}
The expression~\eqref{c33} can be written as the sum~\eqref{act33} over partitions~\eqref{part1}, because
the term corresponding to the partition $\eta_{\scriptscriptstyle \rm I\hspace{-1pt}I}=\{\bar u_j,w\}$ and
$\xi_{\scriptscriptstyle \rm I}=\{w\}$ vanishes due to the presence of the factor ${\sf
f}(\xi_{\scriptscriptstyle \rm I},\eta_{\scriptscriptstyle \rm I\hspace{-1pt}I})$ in the denominator
of~\eqref{act33}.
As above, the other three types of partitions $\xi_{\scriptscriptstyle \rm I}=\eta_{\scriptscriptstyle \rm
I}=\{w\}$; $\xi_{\scriptscriptstyle \rm I}=\{v_i\}$, $\eta_{\scriptscriptstyle \rm I}=\{w\}$;
$\xi_{\scriptscriptstyle \rm I}=\{v_i\}$, $\eta_{\scriptscriptstyle \rm I}=\{u_j\}$ yield exactly the three
terms in~\eqref{c33}, due to~\eqref{reduc}.
The action~\eqref{act33} for $n=1$ is proved.

Before continuing with the action of the lower-triangular monodromy matrix entries $T_{21}(w)$, $T_{32}(w)$
and $T_{31}(w)$ onto the of\/f-shell Bethe vectors, let us run a~check of the
formulae~\eqref{c22-1}, \eqref{c11} and~\eqref{c33}.
It is easy to see that these formulae lead to the Bethe equations when one requires that the vector
$\mathbb{B}^{a,b}(\bar u;\bar v)$ is an eigenvector of the transfer matrix.
Indeed
\begin{gather*}
(T_{11}(w)+T_{22}(w)+T_{33}(w))\mathbb{B}^{a,b}(\bar u;\bar v)=\tau(w;\bar u,\bar v)\mathbb{B}^{a,b}
(\bar u;\bar v),
\end{gather*}
where
\begin{gather*}
\tau(w;\bar u,\bar v)=\lambda_1(w){\sf f}(\bar u,w)+\lambda_2(w){\sf f}(w,\bar u){\sf f}
(\bar v,w)+\lambda_3(w){\sf f}(w,\bar v),
\end{gather*}
provided the Bethe equations
\begin{gather*}
{\sf r}_1(u_j)=\frac{{\sf f}(u_j,\bar u_j)}{{\sf f}(\bar u_j,u_j)}{\sf f}(\bar v,u_j),
\qquad
{\sf r}_3(v_i)=\frac{{\sf f}(\bar v_i,v_i)}{{\sf f}(v_i,\bar v_i)}{\sf f}(v_i,\bar u)
\end{gather*}
are satisf\/ied.
The coef\/f\/icient in front of $\mathbb{B}^{a,b}(\{\bar u_j,w\};\{\bar v_i,w\})$ vanishes due to the
trivial identity
\begin{gather*}
{\sf K}^{(r)}_1(w|v_i){\sf K}^{(r)}_1(v_i|u_j)+{\sf K}^{(l)}_1(u_j|w){\sf K}^{(r)}_1(w|v_i)+{\sf K}^{(l)}
_1(u_j|w){\sf K}^{(l)}_1(v_i|u_j)=0.
\end{gather*}

We now compute the action of the lower-triangular monodromy matrix elements onto of\/f-shell Bethe vectors.
Let us repeat once again the strategy of our calculation, for example, in the case of the action of the
element
\begin{gather*}
T_{21}(w)={k}^+_2(w){\rm E}^+_{12}(w)+{\rm F}^+_{32}(w){k}^+_3(w){\rm E}^+_{13}(w).
\end{gather*}
The calculation of the action in our approach means to normal order the product
\begin{gather}
\label{c21}
T_{21}(w)\cdot{P}^+_f\sk{{F}_2(v_b)\cdots{F}_2(v_1)\cdot{F}_1(u_a)\cdots{F}_1(u_1)}.
\end{gather}
It is done in the context of circular ordering of the Cartan--Weyl or current generators of the quantum
af\/f\/ine algebra $\Uq{3}$ described in subsection~\ref{subal}, and after this ordering one needs to keep
only those terms that belong to the subalgebra $U^+_F$.
According to the presentation~\eqref{us-P2} and the equivalence~\eqref{ac-m-fm}, the r.h.s.\
of~\eqref{c21} can be written as follows
\begin{gather}
\label{c21-1}
{P}^+_f\!\sk{T_{21}(w)\cdot{\cal F}_2(\bar v){\cal F}_1(\bar u)-\sum_{j=1}^a{\sf g}(w,u_j)u_j{\sf f}
(\bar v,u_j)T_{11}(w)\cdot{\cal F}_2(\bar v){\cal F}_1(\bar u_j)\frac{{\sf f}(u_j,\bar u_{>j})}{{\sf f}
(\bar u_{>j},u_j)}}\!,
\end{gather}
where f\/irst we calculate the ordering under projection in~\eqref{c21-1} modulo elements from the ideal
$J$ and then apply projection only to those terms which do not belong to this ideal.
We can simply remove all the elements from the ideal $J$ in~\eqref{c21-1} before taking the projection,
since by def\/inition $J|0\rangle=0$.
Once it is done, we multiply~\eqref{c21} and~\eqref{c21-1} by the product $\beta(\bar u|\bar v){\sf
r}_3(\bar v){\sf f}^{-1}(\bar v,\bar u)$ and act by both of these elements onto right vacuum vector
$|0\rangle$ according to the def\/inition~\eqref{rbv2} to recover the action $T_{21}(w)$ onto
$\mathbb{B}^{a,b}(\bar u;\bar v)$.

Due to the fact that the matrix elements $T_{1\ell}(w)$, $\ell=1,2,3$, act ef\/fectively only on the
f\/irst term in~\eqref{us-P2} we may formally write
\begin{gather*}
T_{1\ell}(w)\cdot{P}^+_f\sk{{\cal F}_2(\bar v)\cdot{\cal F}_1(\bar u)}={P}^+_f\sk{T_{1\ell}(w)\cdot{\cal F}
_2(\bar v)\cdot{\cal F}_1(\bar u)}
\end{gather*}
understanding this equality in the sense described above.
It means that recovering the Bethe vectors in~\eqref{c21-1}, we may f\/irst interchange the projection
${P}^+_f$ and the action of $T_{11}(w)$, then restore the Bethe vector from the projection and f\/inally
use the already calculated action of the monodromy matrix element $T_{11}(w)$ onto $\mathbb{B}^{a,b}(\bar
u;\bar v)$ given by~\eqref{c11}.
This will slightly simplify the whole calculation, although we cannot do the same trick for the calculation
of the remaining matrix elements $T_{ij}(w)$, $i\not=1$.
To calculate the action of these matrix elements onto the of\/f-shell Bethe vectors, we have to use an
explicit expression in terms of the Gauss coordinates and the commutation relations of the Gauss
coordinates with the full currents.

$\bullet$ The action of $T_{21}(w)$.
Taking these rules into account and using~\eqref{ae13-1} and~\eqref{ae12-1} we may calculate
\begin{gather}
T_{21}(w)\mathbb{B}^{a,b}(\bar u;\bar v)=\lambda_2(w)\Bigg(\sum_{j=1}^a{\sf K}^{(r)}_1(w|u_j){\sf r}
_1(u_j)\frac{{\sf f}(w,\bar u_j){\sf f}(\bar u_j,u_j){\sf f}(\bar v,w)}{{\sf f}(\bar v,u_j)}\mathbb{B}
^{a-1,b}(\bar u_j;\bar v)
\nonumber
\\
 \qquad{} +\sum_{\substack{1\leq i\leq b\\1\leq j\leq a}}
{\sf K}^{(r)}_1(w|v_i){\sf K}^{(r)}_1(v_i|u_j){\sf r}_1(u_j)\frac{{\sf f}
(w,\bar u_j){\sf f}(\bar u_j,u_j){\sf f}(\bar v_i,v_i)}{{\sf f}(\bar v,u_j)}\mathbb{B}^{a-1,b}
(\bar u_j;\{\bar v_i,w\})\Bigg)
\nonumber
\\
\qquad
{} +T_{11}(w)\sum_{j=1}^a{\sf K}^{(l)}_1(u_j|w){\sf f}(u_j,\bar u_j)\mathbb{B}^{a-1,b}(\bar u_j;\bar v).
\label{c21-4}
\end{gather}
Then, using~\eqref{c11} the expression~\eqref{c21-4} can be written in the form~\eqref{act21} with a~sum
over partitions of the sets $\bar\eta=\{\bar u,w\}\Rightarrow\{\bar\eta_{\scriptscriptstyle \rm I},
\bar\eta_{\scriptscriptstyle \rm I\hspace{-1pt}I}, \bar\eta_{\scriptscriptstyle \rm
I\hspace{-1pt}I\hspace{-1pt}I}\}$ and $\bar\xi=\{\bar v,w\}\Rightarrow\{\bar\xi_{\scriptscriptstyle \rm
I},\bar\xi_{\scriptscriptstyle \rm I\hspace{-1pt}I}\}$ such that $\#\bar\eta_{\scriptscriptstyle \rm
I}=\#\bar\eta_{\scriptscriptstyle \rm I\hspace{-1pt}I}=\#\bar\xi_{\scriptscriptstyle \rm I}=1$.
Note that in doing so, one possible partition $\bar\xi_{\scriptscriptstyle \rm I}=\{v_i\}$,
$\bar\xi_{\scriptscriptstyle \rm I\hspace{-1pt}I}=\{\bar v_i,w\}$, $\bar\eta_{\scriptscriptstyle \rm
I}=\{w\}$, $\bar\eta_{\scriptscriptstyle \rm I\hspace{-1pt}I}=\{u_j\}$, $\bar\eta_{\scriptscriptstyle \rm
I\hspace{-1pt}I\hspace{-1pt}I}=\{\bar u_j\}$ yields a~zero contribution, due to the factor ${\sf
f}^{-1}(\bar\xi_{\scriptscriptstyle \rm I\hspace{-1pt}I},\bar\eta_{\scriptscriptstyle \rm I})$.
The action~\eqref{act21} for $n=1$ is proved.

$\bullet$ The action of $T_{32}(w)$.
Repeating the same arguments we may present the intermediate result for the action of this matrix element
\begin{gather}
T_{32}(w)\mathbb{B}^{a,b}(\bar u;\bar v)=\lambda_2(w)\left(\sum_{i=1}^b{\sf K}^{(r)}_1(w|v_i){\sf r}
_3(w)\frac{{\sf f}(w,\bar v_i){\sf f}(\bar v_{i},v_i){\sf f}(w,\bar u)}{{\sf f}(w,\bar u)}\mathbb{B}
^{a,b-1}(\bar u;\bar v_i)\right.
\nonumber
\\
\qquad
{} +\sum_{i=1}^b{\sf K}^{(l)}_1(v_i|w){\sf r}_3(v_i)\frac{{\sf f}(v_{i},\bar v_i){\sf f}(\bar v_{i}
,w){\sf f}(w,\bar u)}{{\sf f}(v_i,\bar u)}\mathbb{B}^{a,b-1}(\bar u;\bar v_i)
\nonumber
\\
\left.\qquad{} +\sum_{1\leq i\not=i'\leq b}{\sf K}^{(l)}_1(v_i|w){\sf K}^{(r)}_1(w|v_{i'}){\sf r}
_3(v_i)\frac{{\sf f}(v_{i},\bar v_i){\sf f}(\bar v_{i,i'},v_{i'}){\sf f}(w,\bar u)}{{\sf f}(v_i,\bar u)}
\mathbb{B}^{a,b-1}(\bar u;\{\bar v_{i,i'},w\})\right)  
\nonumber
\\
\qquad
{}+T_{12}(w)\sum_{\substack{1\leq j\leq a\\1\leq i\leq b}}
{\sf K}^{(l)}_1(u_{j}|w){\sf K}^{(l)}_1(v_i|u_j){\sf r}_3(v_i)\frac{{\sf f}(v_{i}
,\bar v_i){\sf f}(u_j,\bar u_j)}{{\sf f}(v_i,\bar u)}\mathbb{B}^{a-1,b-1}(\bar u_{j};\bar v_i).
\label{c32int}
\end{gather}
Using~\eqref{c12-2} we may present~\eqref{c32int} in the form~\eqref{act32} as sum over partitions of the
sets $\bar\eta=\{\bar u,w\}\Rightarrow\{\bar\eta_{\scriptscriptstyle \rm I}, \bar\eta_{\scriptscriptstyle
\rm I\hspace{-1pt}I}\}$ and $\bar\xi=\{\bar v,w\}\Rightarrow\{\bar\xi_{\scriptscriptstyle \rm
I},\bar\xi_{\scriptscriptstyle \rm I\hspace{-1pt}I},\bar\xi_{\scriptscriptstyle \rm
I\hspace{-1pt}I\hspace{-1pt}I}\}$ such that $\#\bar\xi_{\scriptscriptstyle \rm
I}=\#\bar\xi_{\scriptscriptstyle \rm I\hspace{-1pt}I}=\#\bar\eta_{\scriptscriptstyle \rm I}=1$.
The action~\eqref{act32} for $n=1$ is proved.

$\bullet$ The action of $T_{31}(w)$.
The action of the matrix element $T_{31}(w)$ can be calculated analogously.
The intermediate result of this action is
\begin{gather*}
T_{31}(w)\mathbb{B}^{a,b}(\bar u;\bar v)
\\
\qquad
=\!\lambda_2(w)\!\!\left(\sum_{\substack{1\leq j\leq a\\1\leq i\leq b}}\!
{\sf K}^{(r)}_1(v_{i}|u_j){\sf K}^{(r)}_1(w|v_i){\sf r}_1(u_j){\sf r}_3(w)\frac{{\sf f}
(\bar u_j,u_j){\sf f}(w,\bar v_i){\sf f}(\bar v_i,v_i)}{{\sf f}(\bar v,u_j)}\mathbb{B}^{a-1,b-1}
(\bar u_j;\bar v_i)\right.
\\
\qquad{}
+\!\sum_{\substack{1\leq j\leq a\\1\leq i\leq b}}\!
{\sf K}^{(l)}_1(v_{i}|w){\sf K}^{(r)}_1(w|u_{j}){\sf r}_1(u_j){\sf r}
_3(v_i)\frac{{\sf f}(\bar u_j,u_j){\sf f}(w,\bar u_{j}){\sf f}(v_i,\bar v_i){\sf f}(\bar v_{i},w)}{{\sf f}
(v_i,u_j){\sf f}(v_i,\bar u_j){\sf f}(\bar v_i,u_j)}\mathbb{B}^{a-1,b-1}(\bar u_j;\bar v_i)
\\
\qquad{}
+\sum_{\substack{1\leq j\leq a\\1\leq i\not=i'\leq b}}
{\sf K}^{(l)}_1(v_{i}|w){\sf K}^{(r)}_1(v_{i'}|u_{j}){\sf K}^{(r)}_1(w|v_{i'}
){\sf r}_1(u_j){\sf r}_3(v_i)
\\
\left.\qquad{}
\times\frac{{\sf f}(\bar u_j,u_j){\sf f}(w,\bar u_{j}){\sf f}(v_i,\bar v_i){\sf f}(\bar v_{i,i'},v_{i'})}
{{\sf f}(v_i,u_j){\sf f}(v_i,\bar u_j){\sf f}(\bar v_i,u_j)}\mathbb{B}^{a-1,b-1}(\bar u_j;\{\bar v_{i,i'}
,w\})\vphantom{\sum_{\substack{1\leq j\leq a\\1\leq i\leq b}}}\right)
\\
\qquad{}
+T_{11}(w)\sum_{\substack{1\leq j\leq a\\1\leq i\leq b}}
{\sf K}^{(l)}_1(v_{i}|u_j){\sf K}^{(l)}_1(u_{j}|w){\sf r}_3(v_i)\frac{{\sf f}
(u_j,\bar u_j){\sf f}(v_i,\bar v_i)}{{\sf f}(v_i,\bar u)}\mathbb{B}^{a-1,b-1}(\bar u_j;\bar v_i).
\end{gather*}
Using~\eqref{c11} we conclude that the f\/inal result of the action of the monodromy matrix elements
$T_{31}(w)$ can be written in the form~\eqref{act31} as sum over partitions of the sets $\bar\eta=\{\bar
u,w\}\Rightarrow\{\bar\eta_{\scriptscriptstyle \rm I}, \bar\eta_{\scriptscriptstyle \rm I\hspace{-1pt}I},
\bar\eta_{\scriptscriptstyle \rm I\hspace{-1pt}I\hspace{-1pt}I}\}$ and $\bar\xi=\{\bar
v,w\}\Rightarrow\{\bar\xi_{\scriptscriptstyle \rm I},\bar\xi_{\scriptscriptstyle \rm
I\hspace{-1pt}I},\bar\xi_{\scriptscriptstyle \rm I\hspace{-1pt}I\hspace{-1pt}I}\}$ such that
$\#\bar\xi_{\scriptscriptstyle \rm I}=\#\bar\xi_{\scriptscriptstyle \rm
I\hspace{-1pt}I}=\#\bar\eta_{\scriptscriptstyle \rm I}=\#\bar\eta_{\scriptscriptstyle \rm
I\hspace{-1pt}I}=1$.
The action~\eqref{act31} for $n=1$ is proved.

\subsection[The general case $\#\bar w=n$]{The general case $\boldsymbol{\#\bar w=n}$}

We have proved the formulae of the multiple actions~\eqref{act13}--\eqref{act31} for $\#\bar w=1$.
Then the general case $\#\bar w=n$ can be considered via an induction over $n$.
We assume that the equations~\eqref{act13}--\eqref{act31} are valid for $\#\bar w=n-1$ and act
successively: f\/irst by $T_{ij}(\bar w_n)$ and then by $T_{ij}(w_n)$.
The induction for~\eqref{act13} is trivial.
The proofs of the other formulae require the use of lemma~\ref{main-ident}.

Consider, for instance, the multiple action of $T_{23}(\bar w)$.
It is convenient to write~\eqref{act23} in the following form:
\begin{gather}
T_{23}(\bar w_n)\mathbb{B}^{a,b}(\bar u;\bar v)
\nonumber
\\
\qquad{}
=(-q)^{1-n}\lambda_2(\bar w_n)
\sum_{\{\bar w_n,\bar u\}
\Rightarrow\{\bar\eta_{\scriptscriptstyle\rm I},\bar\eta_{\scriptscriptstyle\rm I\hspace{-1pt}I}\}}
{\sf f}(\bar\eta_{\scriptscriptstyle\rm I},\bar\eta_{\scriptscriptstyle\rm I\hspace{-1pt}I})
{\sf K}^{(r)}_{n-1}(\bar w_nq^{-2}|\bar\eta_{\scriptscriptstyle\rm I})
\mathbb{B}^{a,b+n-1}(\bar\eta_{\scriptscriptstyle\rm I\hspace{-1pt}I};\bar\xi).\label{act23-p0}
\end{gather}
Here we have got rid of the poles of ${\sf K}^{(l)}_{n-1}(\bar\eta_{\scriptscriptstyle \rm I}|\bar w_n)$ at
$\eta_i=w_j$ transforming it into ${\sf K}^{(r)}_{n-1}(\bar w_nq^{-2}|\bar\eta_{\scriptscriptstyle \rm I})$
via~\eqref{24}.
Thus, the action of $T_{23}(\bar w_n)$ produces the sum over partitions of the set $\{\bar w_n,\bar u\}$
into subsets $\bar\eta_{\scriptscriptstyle \rm I}$ and $\bar\eta_{\scriptscriptstyle \rm I\hspace{-1pt}I}$.
Applying the operator $T_{23}(w_n)$ to~\eqref{act23-p0} we obtain
\begin{gather}
T_{23}(\bar w)\mathbb{B}^{a,b}(\bar u;\bar v)=(-q)^{-n}\lambda_2(\bar w) \sum_{\{\bar w_n,\bar u\}
\Rightarrow\{\bar\eta_{\scriptscriptstyle\rm I},\bar\eta_{\scriptscriptstyle\rm I\hspace{-1pt}I}\}}{\sf f}
(\bar\eta_{\scriptscriptstyle\rm I},\bar\eta_{\scriptscriptstyle\rm I\hspace{-1pt}I}){\sf K}^{(r)}_{n-1}
(\bar w_nq^{-2}|\bar\eta_{\scriptscriptstyle\rm I})
\nonumber
\\
\qquad{}
\times\sum_{\{w_n,\bar\eta_{\scriptscriptstyle\rm I\hspace{-1pt}I}\}\Rightarrow\{\bar\eta_{\rm i}
,\bar\eta_{\rm ii}\}}{\sf f}(\bar\eta_{\rm i},\bar\eta_{\rm ii}){\sf K}^{(r)}_{1}(w_nq^{-2}|\bar\eta_{\rm i}
)\mathbb{B}^{a,b+n}(\bar\eta_{\rm ii};\bar\xi).\label{act23-p1}
\end{gather}
Here we have an additional sum over partitions of the set $\{w_n,\bar\eta_{\scriptscriptstyle \rm
I\hspace{-1pt}I}\}$ into subsets $\bar\eta_{\rm i}$ and $\bar\eta_{\rm ii}$.
In fact, one can say that we have the sum over partitions of the set $\{\bar w,\bar u\}$ into three subsets
$\bar\eta_{\scriptscriptstyle \rm I}$, $\bar\eta_{\rm i}$, and $\bar\eta_{\rm ii}$ with one additional
constraint $w_n\notin\bar\eta_{\scriptscriptstyle \rm I}$.

Obviously
\begin{gather}
\label{Obv}
{\sf f}(\bar\eta_{\scriptscriptstyle\rm I},\bar\eta_{\scriptscriptstyle\rm I\hspace{-1pt}I})=\frac{{\sf f}
(\bar\eta_{\scriptscriptstyle\rm I},\bar\eta_{\scriptscriptstyle\rm I\hspace{-1pt}I}){\sf f}
(\bar\eta_{\scriptscriptstyle\rm I},w_n)}{{\sf f}(\bar\eta_{\scriptscriptstyle\rm I},w_n)}=\frac{{\sf f}
(\bar\eta_{\scriptscriptstyle\rm I},\bar\eta_{\rm i}){\sf f}(\bar\eta_{\scriptscriptstyle\rm I}
,\bar\eta_{\rm ii})}{{\sf f}(\bar\eta_{\scriptscriptstyle\rm I},w_n)}.
\end{gather}
It is easy to see that the function in the r.h.s.\
of~\eqref{Obv} is a~projector of the product ${\sf f}(\bar\eta_{\scriptscriptstyle \rm
I},\bar\eta_{\scriptscriptstyle \rm I\hspace{-1pt}I})$ onto partitions $\bar\eta_{\scriptscriptstyle \rm
I}$, $\bar\eta_{\rm i}$, and $\bar\eta_{\rm ii}$, such that $w_n\notin\bar\eta_{\scriptscriptstyle \rm I}$:
\begin{gather}
\label{Project}
\frac{{\sf f}(\bar\eta_{\scriptscriptstyle\rm I},\bar\eta_{\rm i}){\sf f}(\bar\eta_{\scriptscriptstyle\rm I}
,\bar\eta_{\rm ii})}{{\sf f}(\bar\eta_{\scriptscriptstyle\rm I},w_n)}=
\begin{cases}
{\sf f}(\bar\eta_{\scriptscriptstyle\rm I},\bar\eta_{\scriptscriptstyle\rm I\hspace{-1pt}I}),
\quad
&
\text{if} \ \
w_n\notin\bar\eta_{\scriptscriptstyle\rm I},
\\
0,
&
\text{if}  \ \
w_n\in\bar\eta_{\scriptscriptstyle\rm I}.
\end{cases}
\end{gather}
Then the sum~\eqref{act23-p1} takes the form
\begin{gather*}
T_{23}(\bar w)\mathbb{B}^{a,b}(\bar u;\bar v)=(-q)^{-n}\lambda_2(\bar w) \sum_{\{\bar w,\bar u\}
\Rightarrow\{\bar\eta_{\scriptscriptstyle\rm I},\bar\eta_{\rm i},\bar\eta_{\rm ii}\}}{\sf K}^{(r)}_{n-1}
(\bar w_nq^{-2}|\bar\eta_{\scriptscriptstyle\rm I}){\sf K}^{(r)}_{1}(w_nq^{-2}|\bar\eta_{\rm i})
\\
\qquad{}
\times\frac{{\sf f}(\bar\eta_{\rm i},\bar\eta_{\rm ii}){\sf f}
(\bar\eta_{\scriptscriptstyle\rm I},\bar\eta_{\rm i}){\sf f}(\bar\eta_{\scriptscriptstyle\rm I}
,\bar\eta_{\rm ii})}{{\sf f}(\bar\eta_{\scriptscriptstyle\rm I},w_n)}\mathbb{B}^{a,b+n}(\bar\eta_{\rm ii}
;\bar\xi).
\end{gather*}
Setting $\{\bar\eta_{\scriptscriptstyle \rm I},\bar\eta_{\rm i}\}=\bar\eta_{0}$ and transforming ${\sf
K}^{(r)}_{1}(w_nq^{-2}|\bar\eta_{\rm i})$ via~\eqref{24} we obtain
\begin{gather}
T_{23}(\bar w)\mathbb{B}^{a,b}(\bar u;\bar v)=(-q)^{1-n}\lambda_2(\bar w) \sum_{\{\bar w,\bar u\}
\Rightarrow\{\bar\eta_{0},\bar\eta_{\rm ii}\}}\frac{{\sf f}(\bar\eta_{0},\bar\eta_{\rm ii})}{{\sf f}
(\bar\eta_{0},w_n)}\mathbb{B}^{a,b+n}(\bar\eta_{\rm ii};\bar\xi)
\nonumber
\\
\hphantom{T_{23}(\bar w)\mathbb{B}^{a,b}(\bar u;\bar v)=}{}
\times\sum_{\bar\eta_{0}\Rightarrow\{\bar\eta_{\scriptscriptstyle\rm I},\bar\eta_{\rm i}\}}{\sf K}^{(l)}_{1}
(\bar\eta_{\rm i}|w_n){\sf K}^{(r)}_{n-1}(\bar w_nq^{-2}|\bar\eta_{\scriptscriptstyle\rm I}){\sf f}
(\bar\eta_{\scriptscriptstyle\rm I},\bar\eta_{\rm i}).
\label{act23-p2}
\end{gather}
The sum over partitions $\bar\eta_{0}\Rightarrow\{\bar\eta_{\scriptscriptstyle \rm I},\bar\eta_{\rm i}\}$
in the last line of~\eqref{act23-p2} can be computed via~\eqref{Klr-ident}, what gives us
\begin{gather*}
T_{23}(\bar w)\mathbb{B}^{a,b}(\bar u;\bar v)=\lambda_2(\bar w) \sum_{\{\bar w,\bar u\}
\Rightarrow\{\bar\eta_{0},\bar\eta_{\rm ii}\}}\frac{{\sf f}(\bar\eta_{0},\bar\eta_{\rm ii}){\sf f}
(\bar w_nq^{-2},\bar\eta_{0})}{{\sf f}(\bar\eta_{0},w_n)}{\sf K}^{(l)}_{n}(\bar\eta_{0}|\bar w)\mathbb{B}
^{a,b+n}(\bar\eta_{\rm ii};\bar\xi).
\end{gather*}
It remains to use ${\sf f}(\bar w_nq^{-2},\bar\eta_{0})={\sf f}^{-1}(\bar\eta_{0},\bar w_n)$, and we arrive
at~\eqref{act23} with $\#\bar w=n$.

All other formulae of multiple actions are proved in exactly the same manner.
Successive action of $T_{ij}(\bar w_n)$ and $T_{ij}(w_n)$ gives a~sum over partitions with constraints.
Introducing appropriate projectors as in~\eqref{Project} we get rid of these constraints.
Then certain sums over partitions can be computed via Lemma~\ref{main-ident}.
The details of these calculations, however, are rather cumbersome, therefore we do not give them here.

\section{Conclusion}

In this paper, we provided the explicit formulae for the monodromy matrix elements acting onto the
of\/f-shell nested Bethe vectors.
Hopefully these formulae will help to calculate the form factors of local operators, in the framework of
the approach developed in~\cite{BelPRS12b}.
As in the case of rational $SU(3)$-symmetric quantum integrable models~\cite{Res86}, it will also lead to
a~formula for the scalar products of the of\/f-shell nested Bethe vectors in quantum integrable models with
${\rm GL}(3)$ trigonometric $R$-matrix.
Indeed, the of\/f-shell Bethe vectors given by formulae~\eqref{rbv2} and~\eqref{lbv2} can be rewritten
through the elements of the monodromy matrix\footnote{Observe that, up to the replacement ${\sf
K}^{(l,r)}_k\to{\sf K}_k$, these formulae have the same structure as the formulae for Bethe vectors in
rational ${\rm GL}(3)$-invariant models.} (see also~\cite{KP-GLN,OPS}):
\begin{gather}
\label{exTbv}
\mathbb{B}^{a,b}(\bar u;\bar v)=\sum\frac{{\sf K}^{(r)}_{k}({\bar v}_{\scriptscriptstyle\rm I}|{\bar u}
_{\scriptscriptstyle\rm I})}{\lambda_2(\bar u_{\scriptscriptstyle\rm I\hspace{-1pt}I})\lambda_2(\bar v)}
\frac{{\sf f}(\bar v_{\scriptscriptstyle\rm I\hspace{-1pt}I},\bar v_{\scriptscriptstyle\rm I}){\sf f}
(\bar u_{\scriptscriptstyle\rm I},\bar u_{\scriptscriptstyle\rm I\hspace{-1pt}I})}{{\sf f}(\bar v,\bar u)}
\,T_{13}({\bar v}_{\scriptscriptstyle\rm I})T_{23}({\bar v}_{\scriptscriptstyle\rm I\hspace{-1pt}I})T_{12}
({\bar u_{\scriptscriptstyle\rm I\hspace{-1pt}I}})|0\rangle,
\\
\label{exTlbv}
\mathbb{C}^{a,b}(\bar u;\bar v)=\sum\frac{{\sf K}^{(l)}_{k}({\bar v}_{\scriptscriptstyle\rm I}|{\bar u}
_{\scriptscriptstyle\rm I})}{\lambda_2(\bar u_{\scriptscriptstyle\rm I\hspace{-1pt}I})\lambda_2(\bar v)}
\frac{{\sf f}(\bar v_{\scriptscriptstyle\rm I\hspace{-1pt}I},\bar v_{\scriptscriptstyle\rm I}){\sf f}
(\bar u_{\scriptscriptstyle\rm I},\bar u_{\scriptscriptstyle\rm I\hspace{-1pt}I})}{{\sf f}(\bar v,\bar u)}
\langle0|\,T_{21}(\bar u_{\scriptscriptstyle\rm I\hspace{-1pt}I})T_{32}
(\bar v_{\scriptscriptstyle\rm I\hspace{-1pt}I})T_{31}(\bar v_{\scriptscriptstyle\rm I}),
\end{gather}
where the sum goes over all partitions of the sets $\bar u\Rightarrow\{\bar u_{\scriptscriptstyle \rm
I},\bar u_{\scriptscriptstyle \rm I\hspace{-1pt}I}\}$ and $\bar v\Rightarrow\{\bar v_{\scriptscriptstyle
\rm I},\bar v_{\scriptscriptstyle \rm I\hspace{-1pt}I}\}$ such that $\#\bar u_{\scriptscriptstyle \rm
I}=\#\bar v_{\scriptscriptstyle \rm I}=k$, $k=0,\dots,\min(a,b)$.
The proof of the formulae~\eqref{exTbv} and~\eqref{exTlbv} will be given elsewhere.
In principle, one can use these formulae to prove the relations~\eqref{act13}--\eqref{act31} using multiple
exchange relations and the properties of the Izergin determinant as it was done in~\cite{BPSR3} for the
${\rm GL}(3)$-invariant integrable models associated with rational $R$-matrix.
However, we showed in this paper that the use of current presentation provides a~simpler way to perform the
calculation.

Combining the explicit presentations~\eqref{exTbv} and~\eqref{exTlbv} with the multiple actions calculated
in the present paper, we can hope to tackle the problem of computing form factors and scalar products.
This strategy was applied successfully to the case of ${\rm GL}(3)$-invariant integrable models associated
with rational $R$-matrix, giving some hope for the trigonometric case.

\appendix

\section{Properties of the Izergin determinant}
\label{apI}

The following properties of the Izergin determinant easily follows from the def\/inition~\eqref{Izer}.

Initial condition:
\begin{gather}
\label{21}
{\sf K}_1(\bar x|\bar y)={\sf g}(x,y).
\end{gather}
Rescaling of the arguments:
\begin{gather*}
{\sf K}_n(\alpha\bar x|\alpha\bar y)=\alpha^{-n}{\sf K}_n(\bar x|\bar y).
\end{gather*}
Reduction:
\begin{gather*}
{\sf K}_n(\bar x,zq^{-2}|\bar y,z)=-\frac{q}{z}{\sf K}_n(\bar x|\bar y)
\qquad
\text{and}
\qquad
{\sf K}_n(\bar x,z|\bar y,zq^2)=-\frac{1}{qz}{\sf K}_n(\bar x|\bar y).
\end{gather*}
Inverse order of arguments:
\begin{gather*}
{\sf K}_n\big(\bar xq^{-2}|\bar y\big)=(-q)^n{\sf f}^{-1}(\bar y,\bar x){\sf K}_n(\bar y|\bar x)
\qquad
\text{and}  
\qquad
{\sf K}_n\big(\bar x|\bar yq^{2}\big)=(-q)^{-n}{\sf f}^{-1}(\bar y,\bar x){\sf K}_n(\bar y|\bar x).
\end{gather*}
Residues in the poles at $x_j=y_k$:
\begin{gather*}
{\sf K}_n(\bar x|\bar y)|_{x_n\to y_n}
={\sf g}(x_n,y_n){\sf f}(y_n,\bar y_n){\sf f}(\bar x_n,x_n){\sf K}_{n-1}(\bar x_n|\bar y_n)+\text{reg},
\end{gather*}
where $\text{reg}$ means regular part.

Using these properties of ${\sf K}_n$ one can easily derive similar properties for its modif\/ications
${\sf K}^{(l,r)}$, in particular,
\begin{gather}
{\sf K}^{(r)}_n(\bar xq^{-2}|\bar y)=(-q)^n{\sf f}^{-1}(\bar y,\bar x){\sf K}^{(l)}_n(\bar y|\bar x)
\qquad
\text{and}\nonumber\\
{\sf K}^{(l)}_n(\bar x|\bar yq^{2})=(-q)^{-n}{\sf f}^{-1}(\bar y,\bar x){\sf K}^{(r)}_n(\bar y|\bar x).\label{24}
\end{gather}

One more important property of ${\sf K}_{n}(\bar x|\bar y)$ is a~summation formula.
\begin{lem}[main lemma] \label{main-ident} Let $\bar\gamma$, $\bar\alpha$ and $\bar\beta$ be three sets of complex variables with
$\#\alpha=m_1$, $\#\beta=m_2$, and $\#\gamma=m_1+m_2$.
Then
\begin{gather}
\label{Sym-Part-old1}
\sum{\sf K}_{m_1}(\bar\gamma_{\scriptscriptstyle\rm I}|\bar\alpha){\sf K}_{m_2}
(\bar\beta|\bar\gamma_{\scriptscriptstyle\rm I\hspace{-1pt}I}){\sf f}
(\bar\gamma_{\scriptscriptstyle\rm I\hspace{-1pt}I},\bar\gamma_{\scriptscriptstyle\rm I})=(-q)^{-m_1}{\sf f}
(\bar\gamma,\bar\alpha){\sf K}_{m_1+m_2}(\{\bar\alpha q^{-2},\bar\beta\}|\bar\gamma).
\end{gather}
The sum is taken with respect to all partitions of the set
$\bar\gamma\Rightarrow\{\bar\gamma_{\scriptscriptstyle \rm I},\bar\gamma_{\scriptscriptstyle \rm
I\hspace{-1pt}I}\}$ with $\#\bar\gamma_{\scriptscriptstyle \rm I}=m_1$ and
$\#\bar\gamma_{\scriptscriptstyle \rm I\hspace{-1pt}I}=m_2$.
Due to~\eqref{24} the equation~\eqref{Sym-Part-old1} can be also written in the form
\begin{gather}
\label{Sym-Part-old2}
\sum{\sf K}_{m_1}(\bar\gamma_{\scriptscriptstyle\rm I}|\bar\alpha){\sf K}_{m_2}
(\bar\beta|\bar\gamma_{\scriptscriptstyle\rm I\hspace{-1pt}I}){\sf f}
(\bar\gamma_{\scriptscriptstyle\rm I\hspace{-1pt}I},\bar\gamma_{\scriptscriptstyle\rm I})=(-q)^{m_2}{\sf f}
(\bar\beta,\bar\gamma){\sf K}_{m_1+m_2}\big(\bar\gamma|\big\{\bar\alpha,\bar\beta q^2\big\}\big).
\end{gather}
\end{lem}

An analog of this lemma was proved in~\cite[Appendix~A]{BelPRS12b}.
The proof of~\eqref{Sym-Part-old1} coincides with the one given in~\cite{BelPRS12b}.

The equations~\eqref{Sym-Part-old1},~\eqref{Sym-Part-old2} yield similar identities involving ${\sf
K}^{(l,r)}$, for instance,
\begin{gather}
\label{Klr-ident}
\sum{\sf K}^{(l)}_{m_1}(\bar\gamma_{\scriptscriptstyle\rm I}|\bar\alpha){\sf K}^{(r)}_{m_2}
(\bar\beta|\bar\gamma_{\scriptscriptstyle\rm I\hspace{-1pt}I}){\sf f}
(\bar\gamma_{\scriptscriptstyle\rm I\hspace{-1pt}I},\bar\gamma_{\scriptscriptstyle\rm I})=(-q)^{m_2}{\sf f}
(\bar\beta,\bar\gamma){\sf K}^{(l)}_{m_1+m_2}\big(\bar\gamma|\big\{\bar\alpha,\bar\beta q^2\big\}\big).
\end{gather}

\subsection*{Acknowledgements} Work of S.P.\ was supported in part by RFBR grant 11-01-00980-a and grant of
Scientif\/ic Foundation of NRU HSE 12-09-0064.
E.R.\ was supported by ANR Project DIADEMS (Programme Blanc ANR SIMI1 2010-BLAN-0120-02).
N.A.S.\ was supported by the Program of RAS Basic Problems of the Nonlinear Dynamics, RFBR-11-01-00440,
SS-4612.2012.1.

\pdfbookmark[1]{References}{ref}
\LastPageEnding

\end{document}